\documentclass[english,aps,10pt,prl, notitlepage, superscriptaddress, longbibliography,  nofootinbib, reprint, floatfix]{revtex4-2}

\usepackage{graphicx} 
\usepackage{bm}
\usepackage[svgnames]{xcolor}
\usepackage{amsmath}
\usepackage{braket}
\usepackage{wasysym}
\usepackage{txfonts}

\newcommand*{\todocite}[1]{\textcolor{purple}{[add citations]}}
\usepackage[normalem]{ulem}

\newcommand*{\heading}[1]{\belowpdfbookmark{#1}{#1}{\bfseries\textit{#1.---}}\ignorespaces}

\let\section\heading

\usepackage{tikz}
\usepackage{circuitikz}
\usetikzlibrary {arrows.meta}
\usetikzlibrary {shapes.arrows}
\usetikzlibrary{decorations.pathmorphing}
\usetikzlibrary{calc}
\usetikzlibrary{positioning}
\usetikzlibrary{fadings}
\tikzfading[name=fade up,
bottom color=transparent!0, top color=transparent!100, middle color=transparent!100]
\usepgflibrary {shadings}
\usetikzlibrary{patterns}

\usepackage{tabularx}

\usepackage{comment}
\iffalse

\else
\excludecomment{gobble}
\fi
\let\heading\section

\everypar=\expandafter{\the\everypar\loosness=-1 }

\usepackage[unicode=true,pdfusetitle,
bookmarks=true,bookmarksnumbered=true,bookmarksopen=true,bookmarksopenlevel=2,
breaklinks=false,pdfborder={0 0 0},backref=false,colorlinks=true]{hyperref}

\begin{document}

\title{Andreev bound-state optoelectronics: absorption spectroscopy \\and microwave-to-optical transduction}
\author{Benjamin Remez}
\affiliation{Joint Quantum Institute,  University of Maryland, College Park, MD 20742, USA.}
\author{Pouyan Ghaemi}
\affiliation{Physics Department, City College of the City University of New York, NY 10031, U.S.A.}
\affiliation{Physics Program, Graduate Center of City University of New York, NY 10031, U.S.A.}
\author{Jay D. Sau}
\affiliation{Condensed Matter Theory Center and Joint Quantum Institute, Department of Physics, University of Maryland, College Park, Maryland 20742, USA}
\author{Mohammad Hafezi}
\affiliation{Joint Quantum Institute, University of Maryland, College Park, MD 20742, USA.}
\date{\today}

\begin{abstract}
    Superconducting weak-link junctions host electron--hole hybridized excitations called Andreev bound states, central to mesoscopic transport and emerging quantum information platforms. Andreev physics has so far been synonymous with the microwave range. However, advancements in superconductor--semiconductor hybrid junctions open the door to the characterization, and manipulation, of Andreev states by light. Here we introduce a model for light--Andreev interaction, with distinct features: Electrons transitioning into Andreev levels can sidestep Pauli exclusion, resulting in two optical absorption resonances separated by twice the bound state energy. One resonance populates the Andreev state and the other empties it, enabling optical control of Andreev qubit parity; pumping both resets the junction and prevents saturation. Given their strong microwave coupling, we show how Andreev bound states can operate as optical--to--microwave transducers with MHz-scale intermode coupling, a key ingredient for heat load management and quantum networking. We illustrate these effects with realistic device parameters. Our results highlight the possibilities in the new field of Andreev optoelectronics.
\end{abstract}
\maketitle
\global\long\def\d{\dagger{}}%
\global\long\def\up{\uparrow{}}%
\global\long\def\down{\downarrow{}}%
\global\long\def\op#1{\hat{#1}^{\vphantom{\dagger}}}%
\global\long\def\dop#1{\hat{#1}^{\d}}%
\global\long\def\vec#1{{\bm{\mathrm{#1}}}}%
\global\long\def\p{\vec{p}}
\global\long\def\k{\vec{k}}%
\global\long\def\hc{\mathrm{H.~c.}}%
\global\long\def\tcbr{\nonumber\\}%

\section{Introduction}
Josephson superconductor--normal-metal--superconductor (SNS) junctions host electron--hole hybridized modes called Andreev bound states (ABSs)  \cite{Andreev1964, Andreev1966, Kulik1969, Bagwell1992}. These discrete subgap states underpin emergent qubit architectures such as  Andreev spin \cite{Chtchelkatchev2003, Hays2021, Lu2025}, level \cite{Zazunov2003, Cheung2024}, and Majorana qubits \cite{Lutchyn2010}. 
Semiconductor weak links allow the ABS spectrum to be tuned electrostatically, forming gatemon qubits \cite{Larsen2015, Casparis2018} and other voltage-controlled devices.
This has been demonstrated with a variety of semiconductors including InAs \cite{deLange2015, Larsen2015,  Casparis2018, Hertel2022, Elfeky2023, Elfeky2025}, Ge \cite{Sagi2024,Hinderling2024,Zheng2024, Kiyooka2025}, InSb \cite{Ke2019, Khan2020, Levajac2023}, and graphene \cite{Nanda2017, Wang2019b,Riechert2025}.  Van der Waals semiconductors like $\text{MoS}_2$ and $\text{WSe}_2$ have also recently been integrated into junctions \cite{Lee2019a,Balgley2025}.

\begin{figure}[h!]
    \centering
    \includegraphics[]{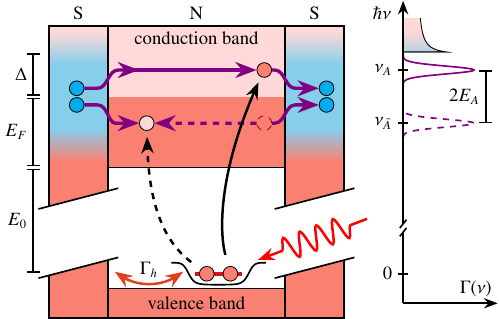}
    \caption{Schematic device band structure and absorption spectrum. A normal semiconductor (N)  is flanked by two superconducting leads (S) with pairing gap $2\Delta$ (cyan) around the doped conduction band Fermi surface $E_F$ (dark pink). Far below, the semiconductor valence band is completely filled. A local potential (black dipper) separates a discrete trapped state from the band continuum. Light (wavy arrow) excites (solid arrow)  the valence electron to the conduction band. At  energies $<\Delta$, this electron (filled circle) hybridizes with a hole below $E_F$ (empty circle) to form a circulating Andreev bound state (purple).
    The hole component allows an \emph{occupied} ABS to absorb a second photoexcited electron (dashed arrow), which we call anomalous absorption. After both transitions, the junction is reset while a Cooper pair (blue circles) was photo-injected into one lead. 
    The circulating electron and hole  have  kinetic energies $E_F\pm E_A$, so the absorption resonances of an empty and occupied ABS are separated by $2E_A$. 
    \label{fig:fig1}}
\end{figure}

Meanwhile, optical control of superconducting circuits is being developed for heat load reduction and remote quantum networking \cite{Lecocq2021,Weaver2024,Warner2025, Arnold2025,vanThiel2025}. 
Though useful superconducting optical devices already exist \cite{Bouscher2017}, they operate either incoherently (e.g. single-photon detectors \cite{Goltsman2001, Natarajan2012, Walsh2021}), or at volt-scale bias incompatible with sub-meV microwave circuits (e.g. Josephson light-emitting diodes \cite{Ozyuzer2007,Asano2009, Recher2010, Hassler2010, Godschalk2011, Mou2015}).
Instead, present interfaces 
\cite{Bochmann2013, Andrews2014, Rueda2016,Jiang2020, Mirhosseini2020, Wu2020}  require additional nonlinear elements for microwave and optical photons to interact \cite{Lambert2020, Han2021}. 
We propose that hybrid SNS junctions, combining semiconductor band gaps in the visible and telecoms range, with Andreev levels that couple strongly to microwaves \cite{Bretheau2013a, Tosi2019, Matute-Canadas2022, Hinderling2024, Elfeky2025}, intrinsically mediate such an interaction.

In this manuscript we develop a minimal model for light--Andreev interaction, where non-trivial effects are already revealed: First we study how ABSs may be characterized by resonant optical absorption, with similarities but also differences to  tunneling spectroscopy \cite{LeSueur2008,Pillet2010,Meschke2011, Bretheau2017, Levajac2023}. An absorbed photon kicks a valence electron into the Andreev level in the conduction band. Strikingly, since the ABSs are hybridized excitations,  an ABS may accept a photoexcited electron whether it is empty \emph{or occupied}, circumventing Pauli blockade in a mechanism we call ``anomalous absorption''. These two transitions are spectrally separated, providing distinct optical knobs to populate and depopulate the junction. This opens the door to manipulating Andreev qubit parity, time-resolved correlation experiments, and shedding light on analogous tangled processes in tunneling spectroscopy \cite{Martin2014, Setiawan2021}.
Lastly, by integrating out the Andreev levels we derive an effective transduction Hamiltonian between microwave and optical photons, showing transduction can be accomplished \emph{in situ} without auxiliary systems.
Our results, which we group under \emph{Andreev optoelectronics}, highlight semiconductor SNS junctions as a platform for combining circuit quantum electrodynamics with the tools and techniques of quantum optics.

\section{Model and geometry}We sketch the proposed device in Fig.~\ref{fig:fig1}. Though the phenomena we describe are general, they are most transparent in semiconductor nanowires \cite{Aguado2020,Badawy2024} where a single transmitting channel makes the system effectively one-dimensional. A semiconducting segment of length $L$ is proximitized by superconducting leads that induce a pairing gap $\Delta(x)$. Impinging photons are absorbed in the semiconductor valence band, exciting electrons into the conduction band where the ABS circulates. This transition is the focus of our manuscript. Resolving this transition requires a discrete valence orbital separated from the band continuum, which can arise naturally from disorder or be engineered using a gated electrostatic potential $V(x)$ \cite[cf. geometries in Refs.][]{Pribiag2013,Higginbotham2015, vanWoerkom2017, Hart2019, Elalaily2024, vanDriel2024}.

Our model Hamiltonian, derived in the Supplementary Material \cite{supnote}, is
\begin{equation} \label{eq:model_H}
    \op{H} \! = (E_0 +E_F)(\op{h}_\up \dop{h}_\up + \op{h}_\down \dop{h}_\down) 
    + E_A(\varphi)(\dop{\gamma}_\up \op{\gamma}_\up - \op{\gamma}_\down \dop{\gamma}_\down) + \op{H}_\mathrm{LMI}
\end{equation}
where  $\dop{\gamma}_\sigma$ ($\dop{h}_\sigma$)  creates an Andreev quasiparticle (trapped valence electron) with spin $\sigma\in\lbrace\up\!,\down\!\!\rbrace$, and $\op{H}_\mathrm{LMI}$ is the light--matter interaction. 
$E_0 + E_F$ is the optical gap from the valence orbital ($-E_0$) to the Fermi level $E_F$.
We adopt the convention of positive-energy ABSs. Their energy $E_A(\varphi)>0$ disperses with the junction phase $\varphi$, and neglecting Coulomb and spin--orbit interactions, it is independent of spin.  The ground state energy $-E_A(\varphi)$ yields the equilibrium Josephson current $I=(2e/\hbar)\partial E_A/\partial\varphi$. 
The discrete orbitals are related to the electron conduction (valence) band field operator $\op{\psi}_{c\sigma (v\sigma)}$ by
\begin{equation} \label{eq:transformation}
    \begin{matrix}
    \op{\psi}_{c\up} (x)  = u(x)\op{\gamma}_\up - v^*(x)\dop{\gamma}_\down\,, \\[5pt] 
    \dop{\psi}_{c\down} (x)  = v(x)\op{\gamma}_\up + u^*(x)\dop{\gamma}_\down\,,
    \end{matrix}
    \quad\text{and} \quad \op{\psi}_{v\sigma}(x)  = f(x)\op{h}_\sigma.
\end{equation}
Here $f(x)$ is the valence trapped orbital, while $u(x)$ and $v(x)$ are the particle and hole amplitudes, respectively, of the ABSs $\dop{\gamma}_{\up,\down}\!=\!\int [ u(x) \dop{\psi}_{c\up,\down}(x)\!\pm\! v(x)\op{\psi}_{c\down,\up}(x)]dx$. $u$ and $v$ are obtained numerically by exactly diagonalizing the Bogoliubov--de-Gennes (BdG)  mean-field Hamiltonian \cite{supnote}. Alongside $\op{\gamma}_\sigma$,  they also have a $\varphi$ dependence which we keep implicit for brevity.

In semiconductor nanowires the optical coupling is maximal for polarization parallel to the wire \cite{Wang2001,Ruda2005}, which couples $s$-like electrons to $p_z$-like $J_z=\pm1/2$ light holes \cite{Los1996, Vezzosi2022}. We model the light--matter interaction by
\begin{equation} \label{eq:LMI-microscopic}
    \op{H}_\mathrm{LMI} = 
    g \sum_\sigma \int_\mathrm{N} \! \! dx\,  \dop{\psi}_{c\sigma}(x)\op{\psi}_{v\sigma}(x) 
    \times \alpha e^{-i\nu t} + \hc, 
\end{equation}
similar to that in  Refs.~\cite{Recher2010, Godschalk2011}. Here $\alpha e^{-i\nu t}$ is a photon field with optical frequency $\hbar\nu \sim E_0 + E_F$ overcoming the bandgap. We consider the field to be the coherent response  $\braket{\op{a}(t)}$ of a single optical mode displaced by an external pump [see Fig.~\ref{fig:transduction_circuit}a], and we neglect its spatial structure. 
That $E_0 \gg \Delta$ permitted  a rotating wave approximation and taking the coupling $g$ as frequency-independent. 
At optical frequencies the metallic leads, whether superconducting or normal, are strongly reflective and transmit/absorb light only weakly \cite{Mattis1958, Rakic1998}. We hence approximate the weak light field in the S regions as zero, signposted by the subscript N in Eq.~\eqref{eq:LMI-microscopic}. Below we discuss  the optical interface further.

$\op{H}_\mathrm{LMI}$ is projected onto the low-energy discrete states by substituting transformation \eqref{eq:transformation}, written
\begin{equation} \label{eq:LMI}
    \op{H}_\mathrm{LMI}\!=\!  
    \left[ \mathcal{D}_e (\dop{\gamma}_\up \op{h}_\up \!+\! \dop{\gamma}_\down \op{h}_\down) + \mathcal{D}_h (\op{\gamma}_\up\op{h}_\down \!-\! \op{\gamma}_\down \op{h}_\up ) \right] \alpha e^{-i\nu t} + \hc
\end{equation}
Here $\mathcal{D}_{e,h}$ are the transition dipole matrix elements given respectively by the ABS electron and hole components, 
\begin{equation} \label{eq:dipoles}
    \mathcal{D}_e  = g \int_\mathrm{N}  u^*(x) f(x) dx,\quad \mathcal{D}_h = g \int_\mathrm{N}  v(x)f(x) dx.
\end{equation}
Though the ABS momentum $k_F$ is mismatched with the zone-center holes, momentum uncertainty $\sim(L+2\xi)^{-1}$ prevents these dipoles from vanishing (see below and the Supplementary \cite{supnote}).
The ABS and valence hole decay rates, which we denote by $\Gamma_A$ and $\Gamma_h$, will set the optical line widths. 
Thanks to the conduction band superconducting gap, the ABS is poisoned on timescales $\Gamma_A^{-1} \apprge 0.1~\mathrm{ms}$ or longer \cite{Janvier2015, Higginbotham2015, Hays2018, Wesdorp2023, Erlandsson2023, MicrosoftAzureQuantum2025}, much slower than the phenomena we will consider.
Conversely, as the valence holes overlap with the leads' Fermi volume [cf. Fig.~\ref{fig:fig1}], we expect them to be comparatively shorter-lived, taking $\Gamma_h \gg \Gamma_A$.

For concrete estimates we take parameters representative of Al-proximitized InAs \cite{Aguado2020,Badawy2024}, with $E_0 \simeq 420$~meV  (wavelength $\lambda\sim3000$~nm), induced gap $\Delta=0.1$~meV, and $E_F=1.2~$meV, corresponding to Fermi momentum $k_F^{-1}\approx35$~nm and coherence length $\xi=\hbar v_F/2\Delta \sim400$~nm. Henceforth we illustrate our results for $L=470$~nm; since $L\sim \xi$ only one Andreev level persists across all $\varphi$. A potential $V(x)= {2(E_F/k_F)Z\delta(x-L_Z)}$ with weak dimensionless strength \cite{Bagwell1992, BTK1982}  $Z = 0.2$ produces a valence state $\approx 0.47 \Delta$ above the band continuum, while retaining near-unity junction transparency  $\tau=1/(1+Z^2)\approx0.96$. 
At position $L_Z = L/4$ it avoids a fine-tuned inversion symmetry. We estimate\footnote{Henceforth factors of $2\pi\hbar$ converting energy to frequency are implicit.} \cite{supnote} $g\sim \sqrt{\alpha_\mathrm{QED}} P /\lambda \sim 28~\text{$\mu$eV}  = 7$~GHz for fine structure constant  $\alpha_\mathrm{QED}\approx1/137$ and $P\sim1~$eVnm the Kane P parameter of InAs \cite{Vurgaftman2001}.
For plotting legibility we will employ a broad $\Gamma_h = 0.02\Delta/\hbar\sim2\pi\times500~$MHz (lifetime ${1/\Gamma_h\sim 0.3}$~ns), but much longer lifetimes can be achieved \cite{Nadj-Perge2010, Hu2012, Petersson2012, vandenBerg2013, Higginbotham2014, Fang2023}.

\section{Anomalous absorption}
The appearance in Eq.~\eqref{eq:LMI} of ``anomalous''\footnote{We say anomalous since it is superficially particle-nonconserving.} terms $\propto \mathcal{D}_h$ is the primary novel feature of this section. 
It shows that valence electrons can be photo-excited into both empty \emph{and occupied} Andreev bound states, thanks to the latter's hybrid nature. This is illustrated in Fig.~\ref{fig:fig1}. An electron transitioning into an occupied ABS depopulates it. 

This process will be evident in the junction's light absorption spectrum:
At cryogenic temperatures, $E_A \sim \Delta \gg k_B T $ so the Andreev sector is empty, $ \braket{\dop{\gamma}_\sigma \op{\gamma}_\sigma } = 0$. However, steady-state absorption rates can be computed in the ground state only if the system can dissipate photons faster than they are absorbed.
The ABSs are long lived, so irradiation can sustain an out-of-equilibrium nonvanishing ABS population. These will support an anomalous optical transition.

We treat absorption within Fermi's Golden Rule,  considering  an  initial Fock state with Andreev occupation $\braket{\dop{\gamma}_\sigma \op{\gamma}_\sigma}=n_{A,\sigma}$  and valence occupation $\braket{\dop{h}_\sigma \op{h}_\sigma}=n_{h,\sigma}$. The absorption spectrum is the transition rate per drive intensity, 
\begin{equation} \label{eq:Gamma}
    \frac{\Gamma(\nu)}{|\alpha|^2} = \frac{\Gamma_h}{\hbar^2} \times \sum_\sigma \Bigg[
    \frac{|\mathcal{D}_e|^2 n_{h,\sigma} (1-n_{A,\sigma})}{(\nu - \nu_A)^2 + \Gamma_h ^2/4}
     + 
    \frac{|\mathcal{D}_h|^2 n_{h,\bar{\sigma}} n_{A,{\sigma}}}{(\nu - \nu_{\bar{A}})^2 + \Gamma_h ^2/4}
    \Bigg]
\end{equation}
where $\bar{\sigma}$ means the spin opposite to $\sigma$. The two terms in Eq.~\eqref{eq:Gamma} correspond to the normal and anomalous transitions, producing resonances at photon energies $\hbar\nu_{A, \bar{A}}=E_0 + E_F \pm E_A$, respectively.  Since $\Gamma_h\gg\Gamma_A$, the former sets the Breit--Wigner resonance line widths.
Eqs.~\eqref{eq:transformation}, \eqref{eq:LMI}, and \eqref{eq:Gamma} easily generalize to multiple ABS and valence levels.

Eq.~\eqref{eq:Gamma} gives instantaneous transition rates, and where $\Gamma(\nu)$ lies with respect to $\Gamma_{A,h}$ will determine the steady state: (i) If $\Gamma \ll \Gamma_A$, $n_{A,\sigma}\approx 0$  remain in equilibrium and the anomalous transition would not be observed. (ii) If $ \Gamma_A \ll \Gamma \ll \Gamma_h$, a nontrivial steady state Andreev population $n_{A,\sigma} \neq 0,1$ can form, so both transition types will be observed provided both resonances $\nu_{A,\bar{A}}$ are pumped. 
(iii) If $\Gamma_h \ll \Gamma$, steady state absorption is limited by the paucity of absorbing valence electrons. Instead re-emission, not captured in Eq.~\eqref{eq:Gamma},  will drive Rabi-like oscillation of electrons between the conduction and valence bands.  Nevertheless,  normal and anomalous transitions involve valence electrons of opposite spin so anomalous transitions can still take place.

That the two transitions involve opposite electron spin, straddle the Fermi level, and reset the ABS, can be understood by considering the electrons' destination: After two transitions, two electrons have been transferred to the conduction band, whereby they pair and enter one of the superconducting leads. Inserting a Cooper pair into the condensate takes twice the chemical potential, and indeed $\nu_A +\nu_{\bar{A}}=2(E_0 + E_F)$ is twice the Fermi energy, measured from the valence orbital.
Weaker two-photon absorption $\propto |\mathcal{D}_e \mathcal{D}_h|^2$, not included in Eq.~\eqref{eq:Gamma}, will appear at frequency $(\nu_A+\nu_{\bar{A}})/2$.

This situation is analogous to tunneling spectroscopy, showing twinned differential conductivity peaks at positive and negative voltage. These are commonly associated, respectively, with the ABS electron ($u^2$) and hole ($v^2$) components \cite{Yazdani1997}. However, this holds only if that state is strongly dissipated by a fermionic bath \cite{Ruby2015, Setiawan2021}. For well-isolated ABSs, as desired for Andreev qubits, the peaks must be equal by particle--hole symmetry, with tunneling $\propto u^2 v^2/(u^2+v^2)$ \cite{Martin2014}. 
This is due to similar ``anomalous'' electron transitions into the occupied ABS. As the probe tip provides electrons of a wide energy range, the normal and anomalous tunneling cannot be controlled individually, the junction never saturates, and the ABS occupation continuously flips. 

In contrast, the two optical absorption peaks map purely to $u$ and  $v$. Since they are spectrally separated, pumping at frequencies $\nu_A$ and $\nu_{\bar{A}}$ can deterministically populate and evacuate the ABS, respectively.\footnote{In this work we neglected spin--orbit and Coulomb interactions. These will make the optical transition spin-selective, and lift the frequency degeneracy between even$\to$odd and odd$\to$even parity transitions, respectively \cite{Wesdorp2023,Kurilovich2024}. This enables the injection of precisely one electron into the Andreev level.} 
This precise parity control is required for the initialization of Andreev level and spin qubits, which operate purely in the even and odd parity sectors, respectively. Microwave methods can bias the slow and stochastic quasiparticle poisoning, which changes the parity only on a millisecond timescale \cite{Wesdorp2023}; optical absorption can occur within nanoseconds, giving considerably faster parity control. 

\begin{figure}[tb]
    \centering
    \includegraphics[width=8.6cm]{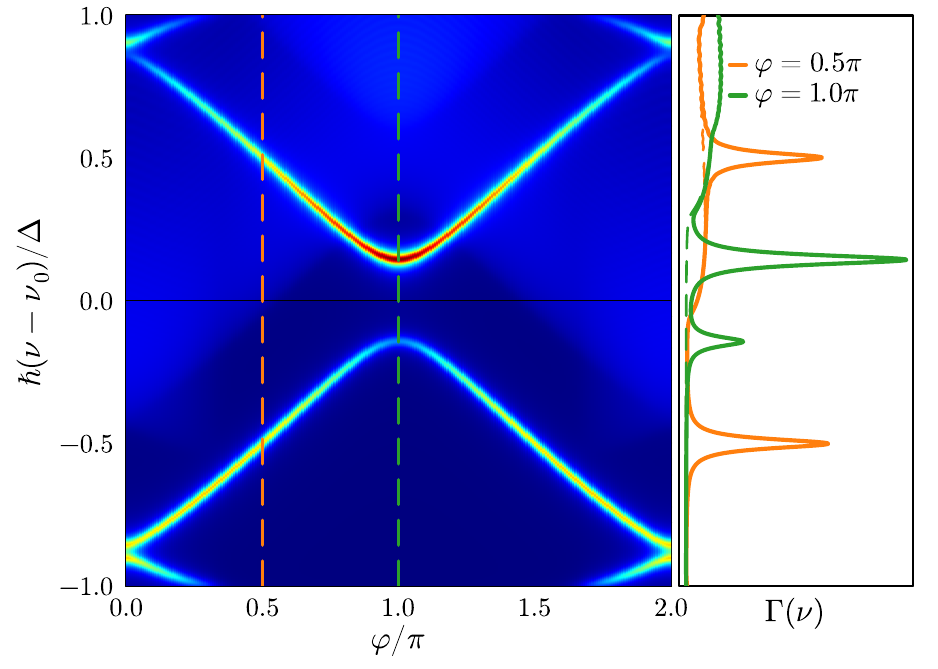}
    \caption{Absorption spectrum versus junction phase $\varphi$. Junction parameters are given in the text; $\hbar\nu_0 = E_0 + E_F$. The right panel shows line cuts at $\varphi/\pi=1/2,1$. The dashed curves indicate the contribution of the valence band continuum.}
    \label{fig:absorption}
\end{figure}

\begin{figure*}
    \centering
    \includegraphics[]{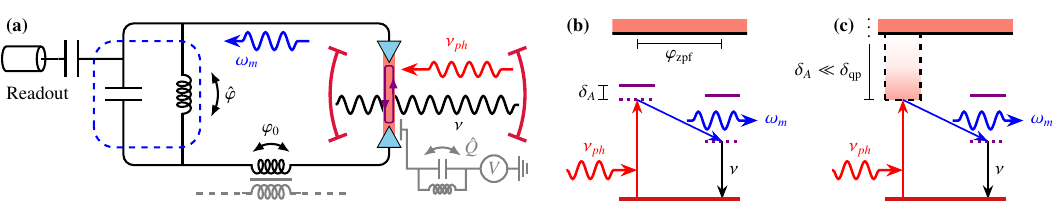}
    \caption{Andreev optical-to-microwave transduction. \textbf{(a)} Circuit diagram. The SNS junction (cyan/pink/cyan) is placed in an optical cavity (red mirrors) and connected to a microwave circuit. External inductive coupling (gray) flux-biases the junction phase to $\varphi \approx \varphi_0$ and allows the ABS persistent current to be detected \cite{Spanton2017,Hart2019}. $\varphi$ is further modulated by fluctuations $\op{\varphi}$  of a microwave resonance $\omega_m$ (blue). The transducer output port is also used to read out the Andreev state by microwave scattering via the ABS-dependent shift of $\omega_m$ (see text). A background pump tone (black wave) bridges the energy gap between microwave and optical photons. 
    ABSs also couples to gate charge fluctuations $\op{Q}$, not pursued in this work. 
    \textbf{(b)} Transduction level diagram, optical$\to$microwave. An optical photon $\nu_{ph}$  (red) virtually excites the ABS (purple, detuning $\delta_A = \nu_{ph}-\nu_A$). Emitting a microwave photon (blue) via the parametric interaction $\partial_\varphi E_A \op{\varphi}\dop{\gamma}\op{\gamma}$ modulates the junction phase by $\sim \varphi_\mathrm{zpf}$. Finally, the pump stimulates de-excitation (black) of the ABS. (c) The optical photon can also excite a virtual above-gap quasiparticle (shaded box, detuning $\delta_\mathrm{qp})$. A similar interaction scatters the quasiparticle into the ABS by emitting a microwave photon. A large $\delta_\mathrm{qp} \apprge  \Delta - E_A \gg \delta_A$ allows us to sum \cite{supnote} over such processes, producing the term $\propto \partial_\varphi\mathcal{D}_e\op{\varphi}$ in Eq.~\eqref{eq:H_microwave}. 
    \label{fig:transduction_circuit}
    } 
\end{figure*}

Fig.~\ref{fig:absorption} depicts the absorption spectrum of our example InAs nanowire. At each $\varphi$ we obtain $E_A$ and $\mathcal{D}_{e,h}$ from exactly diagonalizing the BdG Hamiltonian \cite{supnote}. To display both normal and anomalous transitions, we take $n_{A,\sigma} = 0.5$  and $n_{h,\sigma}=1$. 
The convex and concave curves correspond to the normal and anomalous transitions, respectively, centered around $\hbar\nu_0 = E_0 + E_F$. Their relative prominence  is controlled by the matrix elements $\mathcal{D}_{e,h}(\varphi)$ which disperse with phase.  That $\mathcal{D}_h(\pi) \ll\mathcal{D}_e (\pi)$ is a vestige of inversion symmetry;  an  even $V(x)$ leads to selection rules that fix $D_h(\pi) = 0$ \cite{supnote}.
Similar interference effects can suppress the transitions at high-symmetry $\varphi=0,\pi$ or fine-tuned $k_F L$ and $k_F L_Z$ products. We illustrate other parameter combinations in the Supplementary Material \cite{supnote}.

Here we obtain $\mathcal{D}_e(\pi)\approx 0.055 g\sim 350$~MHz, suggesting all aforementioned absorption regimes are accessible at the single-photon level ($\alpha\apprle 1$) for any sub-GHz linewidth. 
Depending on design, absorption could be measured by the scattering parameters (e.g. $S_{11,12}$) of the optical waveguide, or by resonant near-field response \cite{Knoll1999, Hillenbrand2025}, or most readily, by Andreev microwave spectroscopy \cite[cf.~Fig.~\ref{fig:transduction_circuit}a;][]{Hays2020, Hays2021, MicrosoftAzureQuantum2025}. 
We will describe this method momentarily.

The ABS spectrum deviates from the usual formula \cite{Beenakker1991} $E_A^2=\Delta^2[ 1-\tau\sin^2(\varphi/2)]$  due to the relatively large values of $\Delta/E_F$ and $L/\xi$. The latter also produces a second ABS at $E_{A'}$ near the gap periphery  when  $\varphi\approx0$. As $\varphi$ increases, this ABS merges into the unbound quasiparticle continuum, and its transition dipoles continuously tend to zero as its overlap with the localized hole decreases. Similarly, transitions originating from delocalized valence band continuum states also have only small transition dipole matrix elements. Their weak absorption gives  the faint chevron in Fig.~\ref{fig:absorption}.

\section{Microwave transduction}
Once the optical coupling of the Andreev states has been characterized by absorption spectroscopy, it can be put to work.
So far the ABS wavefunctions depended implicitly on the junction phase $\varphi$. If the junction is embedded in a superconducting circuit, the microwave-range dynamics of  $\varphi$  will couple to the ABSs. This forms the basis for microwave spectroscopy and gate operation in Andreev qubits \cite{Zazunov2003,Bretheau2013a, Cheung2024, deLange2015, Elfeky2025, Hays2018, Hinderling2024, Janvier2015, Matute-Canadas2022, Metzger2021, Tosi2019, vanWoerkom2017, Wesdorp2024}.  Consequently, we propose that the Andreev levels can mediate interactions between microwave and optical photons, namely realizing a coherent microwave--optical photon transducer. 
Though this mechanism is bidirectional, for concreteness we will illustrate optical$\to$microwave downconversion, which addresses the most urgent scalability bottlenecks of cryogenic connectivity and heat loads \cite{Lecocq2021, Arnold2025, Warner2025}, and is more tolerant of optical losses (see below).

We sketch the circuit model in Fig.~\ref{fig:transduction_circuit}a. The phase-dependent circuit ground state energy (to which the ABSs contribute $-E_A$) is least for some $\varphi_0$ that is determined by an external flux bias. 
Promoting the phase variable to quantum operator $\varphi \mapsto \varphi_0 + \op{\varphi}$ and expanding in quantum fluctuations yields the circuit mode $\op{\varphi} = \varphi_\mathrm{zpf}(\dop{b}+\op{b})$ at microwave frequency $\omega_m$. $\varphi_\mathrm{zpf}$ is the mode zero-point phase fluctuations. $\varphi_\mathrm{zpf}\sim 0.1\ll 2\pi$ for 50$\Omega$-matched circuits, allowing us to linearize Hamiltonian \eqref{eq:model_H} in $\op{\varphi}$ \cite{Zazunov2003}. 
Similarly, by promoting $\alpha e^{-i\nu t}\mapsto \alpha e^{-i\nu t}+\op{a}$ we included the optical mode's quantized excitations, with frequency $\nu_{ph}$.
$\nu$, now serving as the pump tone, provides the energy to convert microwave photons to the optical, hence $\nu = \nu_{ph}-\omega_m$.
In the frame rotating with the pump and ABS, the linearized Hamiltonian is
\begin{multline} \label{eq:H_microwave}
   \!\!\!\!\!\!\op{H}   \simeq
    \sum_{\sigma} \partial_\varphi E_A  \op{\varphi} \dop{\gamma}_{\sigma}\op{\gamma}_{\sigma} 
    +  \!\sum_{\sigma}  (\mathcal{D}_e  \!+ \partial_\varphi \mathcal{D}_e \op{\varphi})  (\alpha + \op{a} ) \dop{\gamma}_{\sigma} \op{h}_\sigma + \hc
    \\
     + \sum_\sigma \hbar(\nu_A-\nu) \op{h}_\sigma \dop{h}_\sigma 
     +  \hbar(\nu_{ph} - \nu) \dop{a} \op{a}  + \hbar\omega_m \dop{b}\op{b}.
\end{multline}
Hereafter the  explicit dependence of $E_A, \mathcal{D}_e$, and $\omega_m$ on  $\varphi_0$ is omitted for brevity. 
Below we focus on the ABS vacuum; notwithstanding, the term $\propto \partial_\varphi E_A$ produces an effective ABS-dependent microwave frequency shift $\delta\omega_m\, \dop{\gamma}_\sigma\op{\gamma}_\sigma \dop{b} \op{b}$ that enables fast single-shot Andreev readout \cite{Tosi2019, Hays2020, Metzger2021, Hays2021, MicrosoftAzureQuantum2025}. This presents the most direct means of detecting the optical absorption and parity control of the previous Sections. The ABS persistent current can also be sensed inductively \cite{Spanton2017, Hart2019}.

While Eq.~\eqref{eq:H_microwave} seems like
a trivial differentiation of Eq.~\eqref{eq:model_H}, this Hamiltonian (derived in the Supplementary \cite{supnote}) arises after eliminating higher quasiparticle states, and takes its current form only in a specific gauge and under certain conditions. Namely, we assumed that $E_A$ is isolated from other quasiparticle states (including additional ABSs), and that the optical mode detuning from the Andreev level satisfies $|\mathcal{D}_e| \ll |\hbar\nu_{ph}-\hbar\nu_A|\ll \Delta $ so that the ABS is excited only virtually while the coupling to higher states is perturbatively smaller, respectively. The former condition allows us in this section to dispense with the anomalous terms in Eq.~\eqref{eq:LMI}. Assuming $\omega_m \ll E_A$ allowed us to adiabatically eliminate parity-conserving two-ABS microwave drive terms $\sim \op{\varphi}\dop{\gamma}_\up\dop{\gamma}_\down$. 
Eliminating the higher states may produce an AC Stark shift which we reabsorb into $\nu_A$. These assumptions can be met by tuning $\varphi_0$ away from  $0,\pi$ or reducing the junction transparency somewhat below unity.

Virtual excitations of the Andreev states mediate interactions between the photons. Integrating out the fermions in Eq.~\eqref{eq:H_microwave} and focusing on photon-conserving terms, we obtain the beam splitter-like effective interaction $\op{H}_\mathrm{eff} = \Omega \dop{b}\op{a} + \hc + \dots$ For optical detunings $|\nu-\nu_A|,|\nu_{ph} - \nu_A| \gg \Gamma_h$, the transduction Rabi coupling $\Omega$ takes the form \cite{supnote}
\begin{equation} \label{eq:transduction_Rabi}
    \frac{\hbar \Omega}{2\alpha^* \varphi_\mathrm{zpf}} =      
        \frac{|\mathcal{D}_e|^2 ( \partial_\varphi E_A)/\hbar}{(\nu \!-\! \nu_A)(\nu_{ph} \!-\! \nu_A)} 
        +
        \frac{(\partial_\varphi \mathcal{D}_e^*)\mathcal{D}_e}{\nu_{ph} \!-\! \nu_A} 
        + 
        \frac{\mathcal{D}^*_e (\partial_\varphi \mathcal{D}_e) }{\nu \!-\! \nu_A \!}.
\end{equation}

The processes contributing to $\Omega$ are schematically illustrated  in Fig.~\ref{fig:transduction_circuit}(b,c) and can be understood as follows: The first term in Eq.~\eqref{eq:transduction_Rabi} arises at third order from the sequence (i) absorption of an optical photon virtually excites the ABS, (ii)  emission of a microwave photon by the parametric coupling $\op{\varphi}\dop{\gamma}\op{\gamma}$, and (iii) a pump-induced de-excitation of the ABS. 
The final two terms originate in similar sequences that virtually excite other states, which then scatter into the ABS. Summing over  initial states in the above-gap quasiparticle continuum, as well as doubly microwave-excited ABSs \cite{Bretheau2013a}, produces \cite{supnote} the four-way interaction $\propto{(\partial_\varphi\mathcal{D}_e})\op{\varphi}\op{a}\dop{\gamma}\op{h}$ in Eq.~\eqref{eq:H_microwave}.

\begin{figure}[tb]
    \centering
    \includegraphics[]{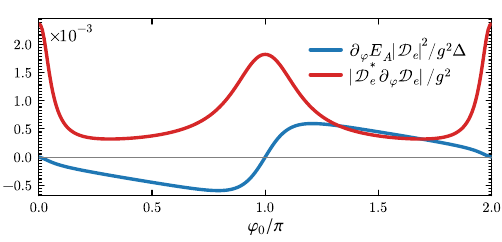}
    \caption{Transduction amplitudes for transitions via a virtual ABS (blue) or the above-gap quasiparticle states (red), corresponding to the numerators in Eq.~\eqref{eq:transduction_Rabi}.
    Note that the former enters Eq.~\eqref{eq:transduction_Rabi} augmented by a factor $\sim \Delta/\hbar\omega_m \gg 1$. Same device parameters as Fig.~\ref{fig:absorption}. }
    \label{fig:transduction_amplitudes}
\end{figure}

We plot the magnitude of these two pathways versus phase bias in Fig.~\ref{fig:transduction_amplitudes}. 
The extra $\partial_\varphi E_A$ factor suggests the first term to be parametrically stronger by $\sim E_A /\hbar\omega_m\gg 1$ at generic bias [e.g. $\varphi_0 = 3\pi/2$]. Conversely, at the  $\varphi_0=0,\pi$ high-symmetry points $\partial_\varphi E_A = 0$, meanwhile  
$\mathcal{D}_e(\varphi)$ has a complex phase that winds rapidly, giving large $\partial_\varphi \mathcal{D}_e \neq 0$ peaks. Together, these permit transduction at any $\varphi_0$. 
For $g\sim 7$~GHz we find $|\mathcal{D}_e|^2 ( \partial_\varphi E_A)\sim 0.6~\text{GHz}^3$ at generic bias and imaginary $\mathcal{D}_e (\partial_\varphi \mathcal{D}^*_e)\sim0.1i~\text{GHz}^2$ at high symmetry, respectively. For $\varphi_\mathrm{zpf}\sim0.1$,  opposite detunings $\nu_{ph}-\nu_A=-(\nu-\nu_A) = \omega_m/2$ [so the last  terms in Eq.~\eqref{eq:transduction_Rabi} add constructively] and $\omega_m/2\pi=5$~GHz, these yield $|\Omega/\alpha| \sim 20~$MHz and $16$~MHz, respectively, or a $\sim 30$~ns photon conversion time.

Importantly, transduction does not require $\Omega\gg \Gamma_h$ \cite{supnote}: Eq.~\eqref{eq:transduction_Rabi} holds if all detunings are $\gg \Gamma_h$.
Meanwhile, the leading photon broadening is $\Gamma(\nu)\ll \Gamma_h$, due to absorption of the far-detuned pump [recall Eq.~\eqref{eq:Gamma}]. It is minimal near $\varphi_0 \sim 0$ [cf. Fig.~\ref{fig:absorption}], $\Gamma(\nu)/|\alpha|^2\sim 0.003\,\Gamma_h$ for our parameters. $\Gamma_h^{-1} \apprge 10~$ns thus yields $\Gamma(\nu)/\Omega \apprle 10^{-3}$. Semiconductor nanowire spin qubits already show longer coherence times, and millisecond relaxation times  \cite{Nadj-Perge2010, Hu2012, Petersson2012, vandenBerg2013, Higginbotham2014, Fang2023}. 
This loss channel also causes pure dephasing, so will not hinder optical projective readout, initialization, etc., in first proofs of concept.

So far we assumed that both the pump and optical mode are closely tuned to the same ABS level $E_A$. Another scheme that could increase $\Omega$ can be devised if say the pump is tuned close to another ABS at some $E_{A'}$ [cf. Fig.~\ref{fig:absorption} and surrounding discussion], thus transducing $\hbar\omega_m \approx |E_{A'} - E_A|$. However, in that case $\op{H}$ will not take the compact form~\eqref{eq:H_microwave}. Additionally, fabricating junctions with length $\apprge \xi$ is challenging (though possible \cite{Park2024}), and the resulting large microwave frequency may be outside experimental capabilities.   
Lastly, microwaves can also be introduced via charge fluctuations in the electrostatic gate \cite[][cf. Fig.~\ref{fig:transduction_circuit}a]{Tosi2019, Metzger2021}. 
Charge fluctuates strongly in low-impedance circuits, so may produce stronger transduction. 
We defer these directions to future work. 

\section{Discussion}
The device we describe here is distinct from previous superconducting optoelectronic elements chiefly in that it couples to \emph{sub-gap, microwave} physics.
The coherence and entanglement of light produced by superconducting \cite{Hayashi2008, Asano2009, Sasakura2011,Hayat2014} and Josephson  \cite{Ozyuzer2007,Recher2010, Godschalk2011} light-emitting diodes is inherited from the superconductor coherent phase. However, in e.g. transmon qubits \cite{Koch2007}, information is encoded not in the phase value, but in its small microwave oscillations. Moreover such sources operate at volt-scale bias \cite{Hayashi2008, Godschalk2011} incompatible with superconducting circuits. More extremely, absorption in superconducting photo-detectors \cite{Goltsman2001, Natarajan2012, Walsh2021} destroys the superconducting state. The Andreev device we propose is therefore the one relevant to superconducting quantum information processing. Our transduction mechanism is also bi-directional, which is not clearly manifest in those systems.

Absorption and transduction lay a gradual road map for the required device integration. 
Optical losses are unimportant for absorption spectroscopy and parity control, requiring only protecting the superconductors from heating by e.g. a mask. Similarly, optically lossy optical$\to$microwave downconversion can still eliminate the significant cryogenic thermal load of microwave control lines \cite{Lecocq2021, Arnold2025, Warner2025}. This enables near-field nanoscopy \cite{Amarie2009, Huth2011, Hillenbrand2025} and other plasmonic methods \cite{Biagioni2012,Dombi2020}, which can confine light below 100~nm, while also trapping the valence states \cite{Dombrowski1999}. Notwithstanding, aluminum thin-films have reflectivity $>$95\% \cite{Rakic1998}, so light spillover in a simple optical microcavity \cite{Vahala2003} may be acceptable for transduction prototypes. These plasmonic optical losses may nonetheless be competitive with current optical heating-limited transducers \cite{Han2021, Xu2024}.
Further on, tapered nanofibers \cite{Stiebeiner2010, Wiedemann2010, He2019a} can provide a low-loss sub-micron optical interface.  Integration will simplify as wider-gap semiconductor ABS platforms, like germanium \cite{Hinderling2024}, mature.

Similarly, engineering efforts can enhance the transition dipoles $\mathcal{D}_{e,h}$.
The comparatively small  $|\mathcal{D}_e|/g\approx0.055$ we found arose from momentum mismatch between the ABS and valence hole, as Eq.~\eqref{eq:dipoles} sieves the $k_F$ component of $f(x)$. The latter has bandwidth $\sim\sqrt{1-\tau}(m_v/m_c)k_F$. For simplicity we took equal band masses $m_v\approx m_c$ as in bulk InAs. Yet heavy holes with $m_v/m_c\sim 16$ \cite[in InAs,][]{Vurgaftman2001} can strongly mix with and weigh down the light holes due to nanowire confinement \cite{Los1996, Lassen2006, Csontos2007, Luo2016,  Csontos2009, Kloeffel2018, Vezzosi2022}. We confirm numerically that a modest doubling of $m_v$ and $Z$ (transparency $0.86$) boost $|\mathcal{D}_e|^2$ by a factor $\apprge 12$. These suggest that through design optimization this coupling can be enhanced.
More consistent hole wavefunctions may be obtained by doping the semiconductor with acceptors. 
Excitonic effects, which dominate semiconductor optical response, could further increase the light--ABS coupling.

\section{Conclusions}
We present a minimal model that couples Andreev bound states to optical-range light, forming the basis for \emph{Andreev optoelectronics}. 
We identify the ``anomalous'' absorption of light by occupied ABSs.
This counteracts junction saturation and allows continuous absorption of bichromatic light.
Absorption can be measured by optical scattering or microwave response.
Unlike tunneling spectroscopy, normal and anomalous absorption are spectrally separated, enabling optical control of Andreev qubit parity.
In addition, Andreev bound states mediate coherent transduction between optical and microwave photons, providing a direct route to optical readout and control of superconducting quantum circuits. 

\.
Looking ahead, pulsed optical driving unlocks a myriad of time-resolved experiments, and may permit the internal dynamics of the junction to be filmed. 
The ABSs also carry the Josephson current, begging the question of whether transport through the junction can be tuned with optical illumination as it can be with microwave irradiation \cite{Gorelik1995,Haxell2023}.  
Finally, beyond interfacing with microwave circuitry, the coupling we describe here may allow optical readout and control of quantum information stored directly in the Andreev levels. 


\section{Acknowledgments} We thank M. Devoret, V. Fatemi, J.~Shabani, K.~Sardashti, and A.~Yacobi  for insightful discussions. J.S. acknowledges support from the Joint Quantum Institute and the Laboratory for Physical Sciences through the Condensed Matter Theory Center. P.G. acknowledges support from NSF DMR 2315063 and 2130544. M.H. and B.R. acknowledge support from the Minta Martin and Simons Foundations and Army Research Office grant W911NF2420240. This research was supported in part by grant NSF PHY-2309135 to the Kavli Institute for Theoretical Physics (KITP).

\bibliography{MyLibrary,\jobname}


\end{document}


\title{Supplementary Material for \emph{Andreev bound-state optoelectronics: absorption spectroscopy \\and microwave-to-optical transduction}}
\author{Benjamin Remez}
\affiliation{Joint Quantum Institute,  University of Maryland, College Park, MD 20742, USA.}
\author{Pouyan Ghaemi}
\affiliation{Physics Department, City College of the City University of New York, NY 10031, U.S.A.}
\affiliation{Physics Program, Graduate Center of City University of New York, NY 10031, U.S.A.}
\author{Jay D. Sau}
\affiliation{Condensed Matter Theory Center and Joint Quantum Institute, Department of Physics, University of Maryland, College Park, Maryland 20742, USA}
\author{Mohammad Hafezi}
\affiliation{Joint Quantum Institute, University of Maryland, College Park, MD 20742, USA.}

\maketitle
\global\long\def\d{\dagger{}}%
\global\long\def\up{\uparrow{}}%
\global\long\def\down{\downarrow{}}%
\global\long\def\op#1{\hat{#1}^{\vphantom{\dagger}}}%
\global\long\def\dop#1{\hat{#1}^{\d}}%
\global\long\def\vec#1{{\bm{\mathrm{#1}}}}%
\global\long\def\p{\vec{p}}
\global\long\def\k{\vec{k}}%
\global\long\def\hc{\mathrm{H.~c.}}%
\global\long\def\tcbr{\nonumber\\}%


\section{Microscopic model}
In this section we describe the microscopic Hamiltonians for the valence and conduction bands.

In the conduction band we use a Bogoliubov--de-Gennes (BdG) mean-field Hamiltonian, usual for describing the SNS system \cite{Kulik1969, Bagwell1992}. Assuming $s$-wave pairing and no spin--orbit coupling, the BdG Hamiltonian takes the form 
$\op{H}_c = \int \dop{\vec{\Psi}} (x) \vec{\mathcal{H}}_c(x) \op{\vec{\Psi}} dx$, where $\op{\vec{\Psi}}=[\op{\psi}_{c\up}\,\,\dop{\psi}_{c\down}]^T$ is a Nambu particle--hole spinor in the conduction band, and the single-particle Hamiltonian density is the two-by-two differential operator
\begin{equation} \label{eq:BdG_H}
    \vec{\mathcal{H}}_{c} (x) = \begin{pmatrix}
      -\frac{\hbar^2}{2m_{c}}\hat{\partial}_x^2 + V(x) -E_F  & \Delta(x) \\
      \Delta^*(x) & E_F +\frac{\hbar^2}{2m_{c}}\hat{\partial}_x^2 - V(x) 
    \end{pmatrix} .
\end{equation} 
Here $E_F$ is the semiconductor Fermi level (measured from the bottom of the conduction band) and $V(x)$ is the electrostatic potential. $\op{H}_c$ will be diagonalized by BdG modes 
$\dop{\gamma}_\lambda = \int [u_\lambda (x) \dop{\psi}_{c\up} (x) + v_\lambda (x) \op{\psi}_{c\down}(x)]dx$
with appropriate particle and hole amplitudes, $u_\lambda (x)$ and $ v_\lambda (x)$, such that  $[\op{H}_c, \dop{\gamma_\lambda}]=E_\lambda\dop{\gamma}_\lambda$. Substituting this into $\op{H}_c$, it can be shown that   $[u_\lambda(x)\,\,v_\lambda(x)]^T$ are an eigenvector of the differential operator $\vec{\mathcal{H}}_c$ with eigenvalue $E_\lambda$. 

Eq.~\eqref{eq:BdG_H} has a particle--hole ({p--h}) anti-symmetry $\op{\Theta} {\vec{\mathcal{H}}}_c \op{\Theta}=-{\vec{\mathcal{H}}}_c$, where $\op \Theta = -i\op{\vec{\eta}}_y \op{\mathcal{K}}$, $\op{\vec{\eta}}_y$ is the second Pauli matrix acting in Nambu space, and $\op{\mathcal{K}}$ is complex conjugation. Therefore, an eigenstate  $[u_\lambda, v_\lambda]$ with eigenvalue $E_\lambda$ has a symmetry partner with amplitudes $[-v^*_\lambda, u^*_\lambda]$ and energy $-E_\lambda$.  Re-labeling $\dop\gamma_\lambda \to \op\gamma_\lambda$ for states with $E_\lambda <0$, each state is now twice degenerate, which corresponds to the SU(2) spin symmetry of $\op{H}$.  This allows us to partition the index $\lambda$ into orbital and particle--hole indices, $\lambda = (\mu,\sigma)$, with  $\mu$ enumerating only the positive-energy eigenstates of $\vec{\mathcal{H}}_c$ and $\sigma = \up,\down$. With this separation we represent
\begin{equation} \label{eq:H_multi_ABS}
    \op{H}_c = \sum_\lambda E_\lambda \dop{\gamma}_\lambda \op{\gamma}_\lambda = \sum_\mu E_\mu (\dop{\gamma}_{\mu\up} \op{\gamma}_{\mu\up} - \op{\gamma}_{\mu\down} \dop{\gamma}_{\mu\down}) 
\end{equation}
and inverse Bogoliubov transformation
\begin{equation}
    \begin{pmatrix}
        \op\psi_{c\up} (x) \\ \dop\psi_{c\down} (x)
    \end{pmatrix} = \sum_\lambda 
    \begin{pmatrix}
        u_\lambda (x) \\ v_\lambda(x)
    \end{pmatrix} \op{\gamma}_\lambda = 
    \sum_\mu \begin{pmatrix}
        u_\lambda (x) \\ v_\lambda(x)
    \end{pmatrix} \op{\gamma}_{\mu\up} + 
    \begin{pmatrix}
        -v_\lambda^* (x) \\ \hphantom{-} u_\lambda^*(x)
    \end{pmatrix} \dop{\gamma}_{\mu\down}.
\end{equation}
In the main text we have restricted these to $\mu = A$ with the lowest $E_A$, representing the one guaranteed Andreev bound state in the junction. 

Now, the valence band Hamiltonian. We assume this band is sufficiently far from the Fermi surface such that any pairing correlations can be neglected, and take a normal  Hamiltonian
\begin{equation}
    \op{H}_v = \sum_\sigma  \psi_{v\sigma}(x) \vec{\mathcal{H}}_v (x) \op\psi_{v\sigma} (x) dx = \sum_\sigma \int \dop\psi_{v\sigma}(x)\left[+\frac{\hbar^2}{2|m_{v}|}\hat{\partial}_x^2 + V(x) - E_F - E_G\right] \op\psi_{v\sigma} (x) dx 
    = 
    \sum_{n,\sigma} (E^{(v)}_n + E_F )  \op{h}_{n\sigma}\dop{h}_{n\sigma}
\end{equation}
where  $E_G$ is the optical bandgap between valence and conduction band extrema, and $|m_v| > 0$ is the inverted valence effective mass. The negative mass causes valence electrons to sense the inverted potential $-V(x)$.
In the last expression we diagonalized this Hamiltonian in terms of valence states indexed by orbital $n$ and spin $\sigma$. Similarly, in the main text we restricted this Hamiltonian to highest valence orbital (smallest $E_n^{(v)}$).

The junction geometry is determined by picking concrete $\Delta(x)$ and $V(x)$.
For the superconducting order parameter $\Delta(x)$ we pick the phase convention \begin{equation}
    \Delta (x) = \begin{cases}
        \Delta e^{i\varphi/2},&  x > L/2 \\
        0,&                     |x|<L/2\\
        \Delta e^{-i\varphi/2},& x < -L/2\,.
    \end{cases}
\end{equation}
whereas for the electrostatic potential we take a single barrier $V = V_0 \delta(x - L_Z)$, $|L_Z| < L/2$.
For the special case that $L_Z = 0$, $\op{H}$ has a combined inversion--conjugation symmetry, see below.

To obtain Figs.~\ref{fig:absorption} and \ref{fig:transduction_amplitudes} in the main text, we numerically diagonalize $\vec{\mathcal{H}}_{c,v}$ by finite differences. The diagonalization volume includes the junction normal region ($|x|<L/2$) and additional 20,000~nm segments of the superconducting leads on either side. This large additional length is required for convergence of the ABSs which decay exponentially into the superconductors, as well as to capture the onset of the vanlence band continuum. We use open boundary conditions. To mitigate numerical noise, we smooth the step (delta) function in $\Delta(x)$ ($V(x)$) into a hyperbolic tangent (gaussian) with characteristic width 10~nm (1~nm), $ \ll $ the Fermi wavelength ($\sim 220~$nm for the parameters used).

We assume for simplicity that $|m_v| = m_c$, which is approximately true for the light hole branch in group III-V semiconductors like InAs \cite{Vurgaftman2001}, and take $m_c = 0.024m_e$, the latter being the bare mass. 
However, this a somewhat stringent assumption: Such semiconductors also host a heavy hole branch with $|m_v|\approx 16 m_c$ \cite{Vurgaftman2001}. Both the trap orbital's binding energy and inverse decay length are $\propto |m_v|$; therefore a heavier hole will separate much more from the valence band continuum, and have momenta much closer to $k_F$, thereby increasing the wavefunction overlap in Eq.~\eqref{eq:dipoles}. Lateral confinement in the nanowire and/or strain lift the bulk light--heavy hole degeneracy, causing them to mix. The hybridized states inherit some of the heavy hole mass, and the resulting valence bands are significantly flattened \cite{Lassen2006, Csontos2007, Kloeffel2018, Vezzosi2022}. The increased effective hole mass will then enhance $|\mathcal{D}_{e,h}|/g$ compared to the values we present in the main text, by a geometry-dependent factor beyond the scope of this work.
For example, we confirmed numerically that doubling the hole mass increases $|\mathcal{D}_e|^2$ by a factor $\approx 4$. A potential valence band inversion \cite{Luo2016} will shift the band edge to finite momentum, further aligning the optical transition in $k$-space.

\subsection{Selection rules at high symmetry}

The combination of high-symmetry phase differences $\varphi = 0, \pi$ and a potential inversion symmetry $V(-x)=V(x)$ will result in additional selection rules that may preclude optical transitions (normal or anomalous) by fixing $\mathcal{D}_{e,h}$ to zero.

We first note that the exact subgap eigenstates of $\op{H}$ are not degenerate. This is natural for generic phase. At $\varphi=0,\pi$, the ABS level crossing \cite{Kulik1969} is protected only by perfect junction transparency, and the crossing is split for generic normal scattering. An even $V(x)$ would lift the degeneracy at $\varphi=\pi$ but not necessarily at $\varphi=0$ \cite{Bagwell1992}. However, that result is derived within the Andreev approximation of no normal reflection at the SN boundaries \cite{Andreev1964, Andreev1966, Kulik1969}, valid when $\Delta/E_F \to 0$. Any finite $\Delta$ will induce normal reflections that will lift the degeneracies at $\varphi=0$ as well. Therefore, the exact eigenstates of $\op{H}$ are not degenerate and hence must be fixed by all symmetries of $\op{H}$.

Now consider the additional symmetry at special values of $\varphi$. Namely, at $\varphi = 0$ ($\pi$) the term proportional to $\Delta$  in $\vec{\mathcal{H}}_c$ can be written $\Delta \vec{\eta}_x$  [$\Delta\mathrm{sgn}(x)i\vec{\eta}_y$], showing that $i\vec{\eta_y}$  ($\vec{\eta}_x$) is a new antisymmetry. Combined with the p--h antisymmetry $i\vec{\eta}_y\mathcal{K}$, it follows that $\mathcal{K}$ and $\mathcal{K}\eta_z$ are symmetries for $\varphi = 0$ and $\varphi = \pi$, respectively.

Additionally, if the electrostatic potential $V(x)$ is even, $\vec{\mathcal{H}}_c$ has an inversion--conjugation symmetry $\mathcal{PK}$, with $\mathcal{P}$ the inversion operation $x\to-x$, valid for all $\varphi$. 
Hence, at fine-tuned $\varphi=0,\pi$ we have $\mathcal{P}$ and $\mathcal{P}\vec{\eta}_z$ as symmetries. That each ABS is not degenerate implies that $u(x)$ and $v(x)$ have the same (opposite) parity at $\varphi = 0$ ($\pi$).
The valence Hamiltonian $\vec{\mathcal{H}}_v$ is also even and therefore the hole wavefunction $f(x)$ will also have definite parity (typically even). 
In total we conclude that at $\varphi=0$ ABSs are characterized as having both $\mathcal{D}_e=\mathcal{D}_h = 0$ or both nonzero, while for $\varphi = \pi$, exactly one of the dipole elements vanishes. 

\subsection{Coupling magnitude}
Here we estimate the magnitude of the light--matter coupling $g$. It is given by the interband matrix element of $\frac{e}{m_e} \op{\vec{A}}\cdot \vec{\op{p}}$ where $\op{\vec{A}}$ is the electromagnetic vector potential and $\op{\vec{p}}$, $m_e$, and $e$ the electron momentum, bare mass, and charge, respectively. The most interesting case is a coupling to a discrete mode, such as that formed by an optical cavity. In that scenario, we estimate
\begin{equation}
    g \approx \sqrt{\frac{\hbar}{2\nu \epsilon_0 V}} \times \frac{e\,p_{cv}}{m_e}{} = \sqrt{\frac{\alpha \lambda}{V}}\times \frac{\hbar p_{cv}}{m_e}\sim \sqrt{\alpha}\frac{\hbar p_{cv}}{m_e \lambda}
\end{equation}
with $\alpha$ the fine structure constant, $\lambda$ the optical wavelength and $p_{cv}$ the interband transition momentum matrix element. $V$ is the optical mode volume, which we take to be $\sim \lambda^3$; in terms of optical focusing, this is still considerably larger than the nanowire volume. InAs has $\lambda\sim$ 3000~nm and $\hbar p_{cv}/m_e\approx 1$~eVnm \cite{Vurgaftman2001}, giving $g\sim 28~\mathrm{\mu eV}$, or $g/2\pi\hbar \sim 7$~GHz. Light polarized parallel to the length of the nanowire is not attenuated by its dielectric constant \cite{Ruda2005}, whereas a high-dielectric optical medium will increase $g\to \sqrt{\epsilon_r}g$, since it allows the mode volume to be shrunk faster than it attenuates the resulting field. For the geometric parameters used in the main text $\mathcal{D}_e\approx0.055\,g$, yielding $\mathcal{D}_e/2\pi\hbar \sim 350~$MHz. As noted above, heavy-hole hybridization may further increase this scale.

\section{Effective transduction Hamiltonian}

In this section we derive the effective microwave transduction Hamiltonian. We slightly rewrite Eq.~\eqref{eq:H_multi_ABS}, including the explicit phase dependence of both the quasiparticle energies and operators,
\begin{equation} \label{eq:Hc_transduction}
    \op{H}_c = \sum_\mu E_\mu(\varphi)\dop{\gamma }_{\mu\sigma}(\varphi)\op{\gamma}_{\mu\sigma}(\varphi) + E_J (\varphi)
\end{equation}
The junction ground state is $E_J(\varphi)=-\sum_\mu E_\mu(\varphi)~+ $ any additional contributions from the microwave environment. It will be minimized at some $\varphi_0$ that can be tuned by an external flux bias. We obtain the microwave Hamiltonian by replacing $\varphi\mapsto \varphi_0 + \op{\varphi}$ and linearizing in the fluctuation $\op{\varphi}$ which we promote to a quantum operator \cite{Zazunov2003}. Neglecting constants and higher nonlinearities, this yields $E_J(\varphi)=\hbar\omega_m \dop{b}\op{b}$, with $\op{b}$ the annihilation operator of a microwave photon at some frequency $\omega_m(\varphi_0)$, and  $\op{\varphi} = \varphi_\mathrm{zpf}(\op{b}+\dop{b})$. Though the BdG Hamiltonian is highly nonlinear in $\varphi$, in low-impedance microwave circuits the zero-point fluctuations $\varphi_\mathrm{zpf}\ll 2\pi $, allowing us nevertheless to linearize it in $\op{\varphi}$. 
The phase dependence of the quasiparticle wavefunctions is lifted to the  fermionic operators, which evolve with phase by
\begin{align}
    \frac{\partial\dop{\gamma}_{\mu\up}}{\partial \varphi} = -i \sum_{\rho} [  \dop{\gamma}_{\rho\up} \mathcal{A}_{\rho\mu} + \op{\gamma}_{\rho\down} \mathcal{B}_{\rho\mu} ],
    \qquad 
    \frac{\partial\dop{\gamma}_{\mu\down}}{\partial \varphi} = -i \sum_{\rho} [ \dop{\gamma}_{\rho\down} \mathcal{A}_{\rho\mu}  -  \op{\gamma}_{\rho\up} \mathcal{B}_{\rho\mu}],
\end{align}
where the matrices $\mathcal{A},\mathcal{B}$ give the Andreev Berry connection
\begin{align} \label{eq:Berry}
    \mathcal{A}_{\rho\mu} = i\int \left[+u^*_\rho(x)\partial_\varphi u_\mu(x) + v^*_\rho(x)\partial_\varphi v_\mu(x)\right] dx, 
    \qquad 
    \mathcal{B}_{\rho\mu} = i\int \left[-v_\rho(x)\partial_\varphi u_\mu(x) + u_\rho(x)\partial_\varphi v_\mu(x)\right] dx,
\end{align}
with $u,v$ evaluated at $\varphi_0$. $\mathcal{A,B}$ are Hermitian and symmetric, respectively.

We now expand Eq.~\eqref{eq:Hc_transduction} by the chain rule. We then add the valence band and light--matter coupling, which do not depend on phase so require no expansion. Moving to the rotating frame described in the main text, we have the Hamiltonian
\begin{align}
    &\op{H}  =  
    \sum_{\mu\sigma} (E_{\mu} - E_{\mu=A}) \dop{\gamma}_{\mu\sigma}\op{\gamma}_{\mu\sigma} 
    +\sum_{\mu\rho\sigma} \Big[ {\delta_{\mu\rho} \partial_\varphi E_\mu
      -i  (E_\mu - E_\rho)  \mathcal{A}_{\rho\mu} }
      \Big]  \dop{\gamma}_{\rho\sigma}\op{\gamma}_{\mu\sigma} \op{\varphi}
    +[ i (E_\mu + E_\rho) \mathcal{B}^*_{\rho\mu}\dop{\gamma}_{\rho\up} \dop{\gamma}_{\mu\down}   + \hc]  \op{\varphi}
    +\hbar\omega_m \dop{b}\op{b}
    \tcbr &
    + \sum_\sigma \hbar (\nu_A - \nu)\op{h}_\sigma\dop{h}_\sigma 
    + \hbar(\nu_{ph}-\nu)\dop{a}\op{a}
    + \sum_{\mu\sigma}[ 
    \mathcal{D}_e^\mu(\varphi_0) \dop{\gamma}_{\mu\sigma} \op{h}_\sigma 
    + (-1)^\sigma \mathcal{D}^\mu_h (\varphi_0)\op{\gamma}_{\mu\sigma}\op{h}_{\bar{\sigma}}] (\alpha + \op{a}) + \hc
\end{align}
We suppress the argument $\varphi_0$ hereafter.  In the last line we generalized Eqs.~\eqref{eq:LMI} and \eqref{eq:dipoles} to the entire quasiparticle spectrum, and defined $(-1)^{\up,\down}=\pm1$. We reiterate that though the dipole matrix elements $D^\mu_{e,h}(\varphi)$ disperse with phase, there is no $\propto\op{\varphi}$ term in the light--matter interaction since the microscopic coupling [cf. Eq.~\eqref{eq:LMI-microscopic} in the main text] does not depend on phase, $\partial_\varphi \op{H}_\mathrm{LMI}=0$.
The term $\propto \mathcal{B}^*$ is well-known as that allowing microwave-driven transitions in the parity-even sector of Andreev level qubits, see e.g. Ref.~\cite{Bretheau2013a}.

Our goal now is to integrate out high energy quasiparticle states to obtain an effective Hamiltonian for one Andreev level at index $\mu=A$. For junction lengths $L\apprle \xi$ and operating point $\varphi_0$ not too close to $0,\pi$, the Andreev levels are well-separated from each other and the continuum by energy of order $\sim \Delta$. The pump tone is tuned to energy conservation, so $\nu = \nu_{ph} - \omega_m$. We now assume the energy hierarchy
\begin{equation}
    \mathcal{D}_e^A \ll |\nu_{ph} -\nu_A|  \ll E_A, \Delta
\end{equation}
The first inequality implies the excitation of ABS $A$ is perturbatively small. The last inequality implies excitations to quasiparticle states $\rho\neq A$ are parametrically weaker; this includes the double-excitation of the $A$ ABS by the $\mathcal{B}$ term. We assume a similar hierarchy for $\omega_m$, implying excitations of more than one photon, either $a$ or $b$, is suppressed. Therefore, our low-energy manifold includes the ABS vacuum and at most one excitation in the $A$ state. Note the latter assumption requires $\varphi_0$ not too close to $\pi$, or alternatively a transparency sufficiently below unity. 

At lowest order, integrating out higher quasiparticle states will generate several kinds of terms: (i) a vacuum Stark shift  renormalizing $E_A$, which we absorb into a redefinition of $E_A$; (ii) four-wave mixing terms like $\sim \dop{\gamma}_A\op{\gamma}_A\hat{b}^{(\d)}\hat{b}^{(\d)}$ which will be weak perturbations by our assumption of large $\omega_m$ and small $\varphi_\mathrm{zpf}$; and (iii) a phase-dependent renormalization of $\mathcal{D}_e^A$. Focusing on the latter, we obtain the term
\begin{align}
    \op{H}_\mathrm{eff} & \ni \sum_{\rho\neq A} [ -i  (E_\rho - E_A)  \mathcal{A}_{A\rho} ] \dop{\gamma}_{A\sigma}  \op{\varphi}\left(\frac{\alpha \mathcal{D}_e^{\,\rho}}{\hbar\nu-(E_0+E_F+E_\rho)} + \frac{\op{a}\mathcal{D}_e^{\,\rho}}{\hbar\nu_{ph}-(E_0+E_F+E_\rho)} \right) \op{h}_\sigma  \tcbr
    & - \sum_\rho \mathcal{D}_h^{\,\rho} \op{h}_\sigma \left(\frac{\dop{b}}{-E_A-E_\rho-\omega_m} + \frac{\op{b}}{-E_A-E_\rho+\omega_m}\right) \varphi_\mathrm{zpf}[ i (E_A + E_\rho) \mathcal{B}^*_{A\rho}]\dop{\gamma}_{A\sigma} (\alpha+\op{a})\tcbr 
    & \approx
     \sum_{\rho\neq A} [ i  (E_\rho - E_A)  \mathcal{A}_{A\rho} ] \dop{\gamma}_{A\sigma}  \op{\varphi}\left(\frac{(\alpha + \op{a})\mathcal{D}_e^{\,\rho}}{E_A - E_\rho} \right) \op{h}_\sigma  
     + \sum_\rho \mathcal{D}_h^{\,\rho} \op{h}_\sigma \left(\frac{(\op{b} + \dop{b})\varphi_\mathrm{zpf}}{E_A+E_\rho} \right) [ i (E_A + E_\rho) \mathcal{B}^*_{A\rho}]\dop{\gamma}_{A\sigma} (\alpha+\op{a})
     \tcbr
     & =  \dop{\gamma}_{A\sigma} \op{h}_\sigma \op{\varphi} (\alpha+\op{a}) \left[i\sum_\rho (\mathcal{A}_{A\rho}\mathcal{D}_e^{\,\rho} - \mathcal{B}_{A\rho}^*\mathcal{D}_h^{\,\rho} ) - i \mathcal{A}_{AA}\mathcal{D}_e^{A}\right] 
     =
     \dop{\gamma}_{A\sigma} \op{h}_\sigma \op{\varphi} (\alpha+\op{a}) \left[\partial_\varphi \mathcal{D}_e^A - i \mathcal{A}_{AA}\mathcal{D}_e^{A}\right]
\end{align}
Here in the third line we employed our assumptions that $\hbar\nu,\hbar\nu_{ph}\approx E_0 +E_F + E_A$ up to a detuning much smaller than $(E_A - E_{\rho\neq A})\sim \Delta$, and $\omega_m \ll E_{\rho\neq A}\sim\Delta$. 
Furthermore, in the last line we used the identity
\begin{equation}
    i\sum_\rho (\mathcal{A}_{\mu\rho}\mathcal{D}_e^{\,\rho} - \mathcal{B}_{\mu\rho}^*\mathcal{D}_h^{\,\rho} ) = \frac{\partial}{\partial \varphi}\mathcal{D}_e^\mu,
\end{equation}
that is guaranteed by gauge invariance and can be verified by substituting Eqs.~\eqref{eq:dipoles} and \eqref{eq:Berry}.
The resulting last bracket is gauge invariant, but we may specialize to the so-called parallel transport gauge, in which the diagonal connections are fixed to zero, $\mathcal{A}_{\mu\mu} = 0$. 
Dropping the $A$ index of the remaining ABS,  this gives the compact effective term
\begin{equation}
    \op{H}_\mathrm{eff} \ni \sum_\sigma  \dop{\gamma}_{\sigma} \op{h}_\sigma (\partial_\varphi \mathcal{D}_e)\op{\varphi} (\alpha+\op{a})
\end{equation}
which we incorporate into Eq.~\eqref{eq:H_microwave} in the main text.

\subsection{Tranduction matrix elements}

To find the effective optical--microwave Hamiltonian, we now integrate out the fermion excitations of the junction (ABSs and holes). Our goal is the effective photon Rabi coupling, corresponding to the photon conversion rate. Eq.~\eqref{eq:H_microwave} separates into independent Hamiltonians corresponding to fermion spin $\sigma=\up,\down$; the virtual states used for conversion are far-detuned, and thus their occupation is parametrically small, so we may consider one spin in isolation and then double the resulting coupling.

The elimination of the fermions is most straightforwardly done with Green's function methods. The complete Green's function of the system is a matrix $G_{mn}(t)=-i\Theta(t)\braket{m(t)|n}$ between any two (many-body) states $\ket{m},\ket{n}$. It is found in the frequency domain from its inverse,
\begin{equation}
    \vec{G}^{-1}(\omega) = \omega - \op{H}/\hbar-\hat{\Sigma}_h = \vec{G}^{-1}_0(\omega) - \hat{\Sigma}^+ - \hat{\Sigma}^- - \hat{\Sigma}^\varphi \label{eq:Greens_function}
\end{equation}
where $\op{H}$ is the Hamiltonian of Eq.~\eqref{eq:H_microwave} (restricted to one spin species). The hole self-energy $\hat{\Sigma}_h=-i(\Gamma_h/2)\op{h}\dop{h}$ accounts for the finite lifetime of the valence hole, and arises from integrating out a larger fermionic bath to which the hole can migrate. This framework holds for sufficiently short times such that number of particles lost to the bath is  $\ll1$, which we verify below. We henceforth reabsorb it in $\op{H}$ by replacing $\nu_A \to \nu_A - i\Gamma_h/2$. 
We then separated the contributions to $\vec{G}^{-1}$ to a diagonal noninteracting part and interaction self-energies,
\begin{gather}
    \vec{G}_0^{-1}(\omega)   =\omega - (\nu_A-\nu-i\Gamma_h/2) \op{h} \dop{h}  -  (\nu_{ph} - \nu) \dop{a} \op{a}  - \omega_m \dop{b}\op{b}; \\  
    \hbar\hat{\Sigma}^-  =  (\mathcal{D}_e  \!+ \partial_\varphi \mathcal{D}_e \op{\varphi})  (\alpha + \op{a} ) \dop{\gamma} \op{h};  \qquad
    \hat{\Sigma}^+ = (\hat{\Sigma}^-)^\d  ; \qquad 
    \hbar\hat{\Sigma}^\varphi  = \partial_\varphi E_A  \op{\varphi} \dop{\gamma}\op{\gamma}.
\end{gather}
A substitution shows that $\vec{G}$ satisfies the Dyson equation
\begin{equation}
    \vec{G}=\vec{G}_0  + \vec{G}_0 (\hat{\Sigma}^+ + \hat{\Sigma}^- + \hat{\Sigma}^\varphi)\vec{G}.
\end{equation}

The column and row indices of $\vec{G}$ can be partitioned into four blocks, corresponding to the 4 possible fermion states. Our goal is to find the components of $\vec{G}$ in the block corresponding to $\dop{\gamma}\op{\gamma}=\op{h}\dop{h}=0$. For this, we write $\vec{G}=\vec{G}_1 + \vec{G}_2$, with $\vec{G}_{1,2}$ the block-diagonal and block-off-diagonal components of $\vec{G}$, respectively. 
Noting $\hat{\Sigma}^\varphi$ is block-diagonal and $\hat{\Sigma}^\pm$ are block-off-diagonal, this lets us separate the Dyson relation into its block-diagonal and off-diagonal components,
\begin{align}
    \vec{G}_1  = \vec{G}_0 + \vec{G}_0 \hat{\Sigma}^\varphi\vec{G}_1 + \vec{G}_0(\hat{\Sigma}^++\hat{\Sigma}^-)\vec{G}_2\,, \qquad\qquad
    \vec{G}_2   = \vec{G}_0 (\hat{\Sigma}^++\hat{\Sigma}^-) \vec{G}_1 + \vec{G}_0 \hat{\Sigma}^\varphi\vec{G}_2  
\end{align}
Eliminating $\vec{G}_2$, we have 
\begin{equation}
    \vec{G}_1 = \vec{G}_0 +  \vec{G}_0 \hat{\Sigma}^\varphi\vec{G}_1 + \vec{G}_0 (\hat{\Sigma}^++\hat{\Sigma}^-)[1-\vec{G}_0 \hat{\Sigma}^\varphi]^{-1} \vec{G}_0 (\hat{\Sigma}^++\hat{\Sigma}^-) \vec{G}_1 =: \vec{G}_0 + \vec{G}_0\hat{\Sigma}_\mathrm{eff}\vec{G}_1
\end{equation}
from which we read off the block-diagonal self-energy
\begin{align}
    \hat{\Sigma}_\mathrm{eff}(\omega) &=  \hat{\Sigma}^\varphi+  (\hat{\Sigma}^++\hat{\Sigma}^-)[1-\vec{G}_0 \hat{\Sigma}^\varphi]^{-1} \vec{G}_0 (\hat{\Sigma}^++\hat{\Sigma}^-) 
    =  \hat{\Sigma}^\varphi+  (\hat{\Sigma}^++\hat{\Sigma}^-)[1+\vec{G}_0 \hat{\Sigma}^\varphi] \vec{G}_0  (\hat{\Sigma}^++\hat{\Sigma}^-) + \mathcal{O}(\hat{\varphi}^2) \tcbr
    & =  \hat{\Sigma}^+\vec{G}_0\hat{\Sigma}^-  + \hat{\Sigma}^+\vec{G}_0 \hat{\Sigma}^\varphi \vec{G}_0 \hat{\Sigma}^-
     + \text{terms vanishing in the fermion ground  state block.}
     \label{eq:self_energy}
\end{align}
In the first line we expanded perturbatively in $\hat{\Sigma}^\varphi$ to focus on single-photon processes. This requires $\varphi_\mathrm{zpf}(\partial_\varphi E_A)/\hbar\omega_m < 1$ (the inequality need not be strong); for high transparency this reads to $\hbar \omega_m \apprge \varphi_\mathrm{zpf}\Delta/2\pi \sim \Delta/60$, holding readily for $\omega_m/2\pi \apprge 1$~GHz.

This self-energy constitutes our effective Hamiltonian, and  the Rabi coupling we seek is $\Omega=\hbar\braket{b|\op{\Sigma}_\mathrm{eff}(\omega)|a}$, yielding
\begin{align}
    &\hbar\Omega(\omega) =   
     \bra{\gamma^\d hb}(\mathcal{D}^*_e  \!+ \op{\varphi} \partial_\varphi  \mathcal{D}^*_e)  (\alpha^* + \dop{a} )\times (1+\vec{G}_0 \hat{\Sigma}^\varphi ) \vec{G}_0\times
    (\mathcal{D}_e  \!+ \op{\varphi}\partial_\varphi \mathcal{D}_e ) (\alpha + \op{a} ) \ket{\gamma^\d ha}\tcbr
    &= \alpha^* \varphi_\mathrm{zpf} \left[|\mathcal{D}^*_e|^2 (\partial_\varphi E_A)\braket{\gamma^\d hb|\vec{G}_0|b\gamma^\d h} \braket{\gamma^\d h|\vec{G}_0|\gamma^\d h} 
    +\mathcal{D}^*_e(\partial_\varphi \mathcal{D}_e ) \braket{\gamma^\d hb |\vec{G}_0 | \gamma^\d h b}
    +\mathcal{D}_e(\partial_\varphi  \mathcal{D}^*_e) \braket{\gamma^\d h|\vec{G}_0|\gamma^\d h}\right] \tcbr
    &=
    \alpha^* \varphi_\mathrm{zpf} \left[
   \frac{ |\mathcal{D}_e|^2 (\partial_\varphi E_A)}{(\omega-(\nu_A-\nu+\omega_m - i\Gamma_h/2))(\omega-(\nu_A-\nu-i\Gamma_h/2))} 
    +\frac{\mathcal{D}^*_e(\partial_\varphi \mathcal{D}_e )}{\omega-(\nu_A-\nu+\omega_m - i\Gamma_h/2)}
    +\frac{\mathcal{D}_e(\partial_\varphi  \mathcal{D}^*_e)}{\omega-(\nu_A-\nu-i\Gamma_h/2)}\right]. \label{eq:full_Rabi}
\end{align}
$\Omega(\omega)$ is to be evaluated at the dressed photon mode frequencies. Under the condition $\Omega \ll \omega_m$, which we verify in the main text, we can safely substitute $\omega \approx\omega_m$. Plugging in energy conservation $\nu_{ph}=\nu+\omega_m$, taking the case that $|\nu_A - \nu|,|\nu_A-\nu_{ph}|\gg \Gamma_h/2$, and doubling for both spin species, we arrive at Eq.~\eqref{eq:transduction_Rabi} of the main text. 

Note that coherent transduction is possible even for $\Gamma_h$ much larger than  $\Omega$ or the microwave photon linewidth. From Eq.~\eqref{eq:full_Rabi} it is evident that for detunings $\delta = |\nu-\nu_A|,|\nu_{ph}-\nu_A|$ much larger than $\Gamma_h/2$, the magnitude of $\Omega$ is not adversely affected. This can be understood from the virtual state picture: the population of the virtual ABS-hole state is $\sim \Omega/\delta$, leading to a probability leakage $T \Omega\Gamma_h/\delta$ after time $T$. A $\pi$-gate rotating a photon between modes is realized if the classical drive is on for $T=\pi/\Omega$, so during operation the leakage is $\sim \Gamma_h/\delta \ll 1$ for large detunings, and justifies the self-energy treatment in Eq.~\eqref{eq:Greens_function}

The finite hole lifetime leads to some loss and decoherence, which we estimate from the diagonal elements of $\hat{\Sigma}_\mathrm{eff}$. In the fermion vacuum block, the dominant (leading in $\varphi_\mathrm{zpf}$ and photon occupations) contribution comes from the first term of Eq.~\eqref{eq:self_energy},
\begin{equation}
    \hat{\Sigma}_{\mathrm{eff},mn}(\omega) \ni  \frac{|\mathcal{D}_e|^2 |\alpha|^2}{\hbar^2}\braket{\gamma^\d h| \vec{G}_0(\omega)| \gamma^\d h}= \frac{|\mathcal{D}_e|^2 |\alpha|^2 \delta_{mn}/\hbar^2}{\omega -E_{m}/\hbar - (\nu_A-\nu-i\Gamma_h/2)}
\end{equation}
At the frequency of both photon modes ($\omega=\nu_{ph}-\nu=\omega_m$), this has the imaginary part 
\begin{equation}
    -2 \mathrm{Im}[\hat{\Sigma}_{\mathrm{eff}}(\omega_m)] = 
    \frac{|\mathcal{D}_e|^2 |\alpha|^2/\hbar^2}{(\nu-\nu_A)^2 + \Gamma_h^2/4} \times  \Gamma_h \approx   \frac{|\mathcal{D}_e|^2 |\alpha|^2}{(\hbar\nu-E_A)^2} \times  \Gamma_h 
\end{equation}
which we recognize (after doubling for spin) as precisely the absorption rate $\Gamma(\nu)$ [see Eq.~\eqref{eq:Gamma} of the main text], with $n_A = 1 - n_h = 0$. This signals that the dominant loss mechanism is real (as opposed to virtual) absorption of a quanta from the classical pump, populating the ABS. In the last equality we took $|\nu-\nu_A|\gg \Gamma_h/2$; by assumption the detuning is also $\gg |\mathcal{D}_e|$, showing that $\Gamma(\nu)\ll\Gamma_h$. We therefore emphasize that that the photons are \emph{not} broadened by the entire hole linewidth, only inheriting a small fraction. For the parameters used in the main text, and operating around $\varphi_0\approx 0$, we find $\Gamma(\nu) \sim 10^{-3}\Gamma_h$. For hole lifetimes $\Gamma_h^{-1} \apprge 1 $~ns this allows for $\Gamma(\nu)/\Omega\ll 1$, such that this loss is small during one transduction cycle.

Furthermore, this absorption does not affect the quantized photon modes, so does not represent true photon loss.  Rather, it leads only to dephasing between the incoming and outgoing photons. For certain applications, such as optical projective readout of the microwave resonator, this is not an impediment. 

\section{Additional parameter combinations}

The results we present in the main text are qualitatively unchanged as the junction parameters are varied. However, quantitative magnitude of our results is somewhat sensitive to these parameter values. In this Appendix we present this dependence for completion. 

The most tunable quantity is $\mathcal{D}_{e,h}/g$, which is determined by the wavefunction overlap between the trapped valence hole and the Andreev electron and hole components. Since the ABS and valence states have disparate momentum contents, their overlap can depend interferometrically on geometric details. Therefore,  the most sensitive knobs for influencing $\mathcal{D}_{e,h}$ are the junction geometric length $L$ and position of the trapping potential $L_Z\in[-L/2, L/2]$ within it.

In Figs.~\ref{fig:extra_params_1} and \ref{fig:extra_params_2} we present the absorption spectrum and transduction amplitudes, counterpart to Figs.~\ref{fig:absorption} and \ref{fig:transduction_amplitudes}, for various combinations of $L$ and $L_Z$ (measured as a fraction of $L$). As we see, the behavior we describe is the text is quite general for most parameter combinations. To reiterate these are: The absorption spectrum has two resonance lines that stand out from valence band-edge transitions, and either transduction amplitude is nonzero for all junction phase biases.

However, quantitative differences can also be gleamed. $L_Z = 0$ (left column of panels) gives a spurious inversion symmetry (see previous sections) which causes $\mathcal{D}_h(\varphi=\pi)$ to vanish, as well as the transduction coefficients $\propto \partial_\varphi \mathcal{D}_e(\varphi=0,\pi)$.
Furthermore, the length dependence can lead to  strong suppression of the dipole transition at fine-tuned points. For example, with $L=360$~nm and $L_Z = 0$, this is evidenced in a $\sim 10^{-4}$ suppression of the transduction coefficients; as a result the resonant absorption lines are drowned by transitions from the valence band continuum. 

Similarly, also at $L=360$~nm but $L=L_Z/10$, we see an atypical inversion of the relative absorption intensities: The anomalous absorption line (concave curve) is brighter than the normal absorption line (convex curve).

Finally, the contribution of the valence band continuum also changes somewhat: While for some combinations (e.g. $L=580$~nm) this background contribution is roughly constant, at other combinations (e.g. $L=690$~nm, and especially in the $L_Z = L/4$ case) there is strong enhancement from the continuum edge. Yet in all these cases the resonant lines generally remain resolvable.

\begin{figure*}[tb]
    \centering
    \includegraphics[width=5.85cm]{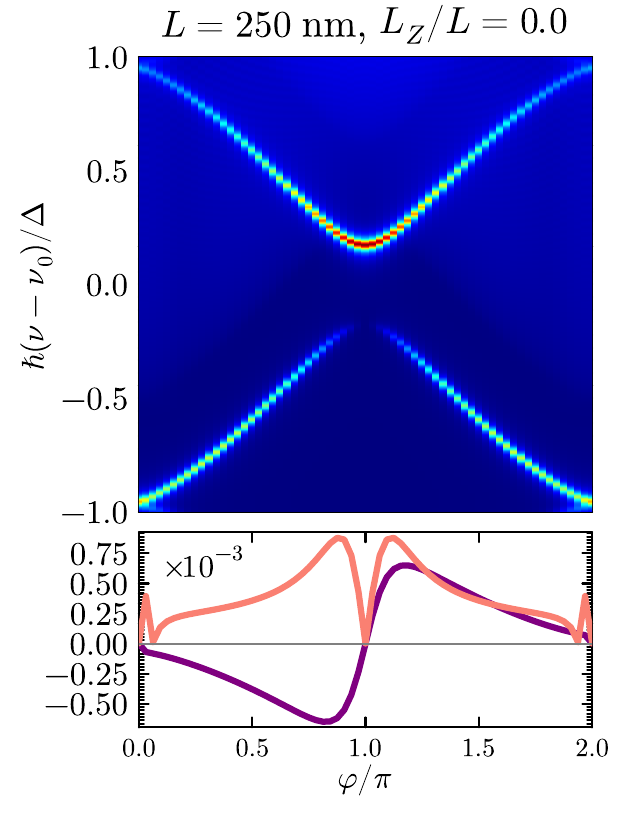}
    \includegraphics[width=5.85cm]{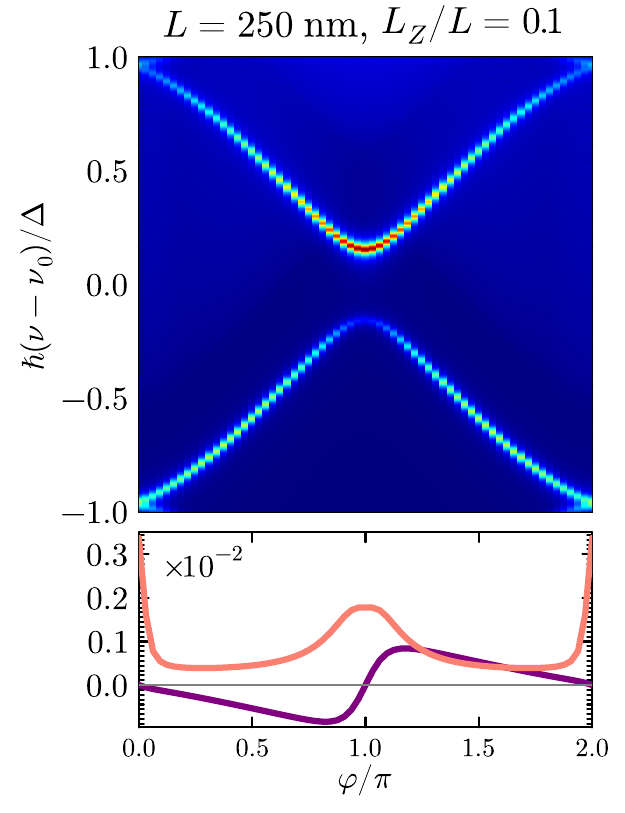}
    \includegraphics[width=5.85cm]{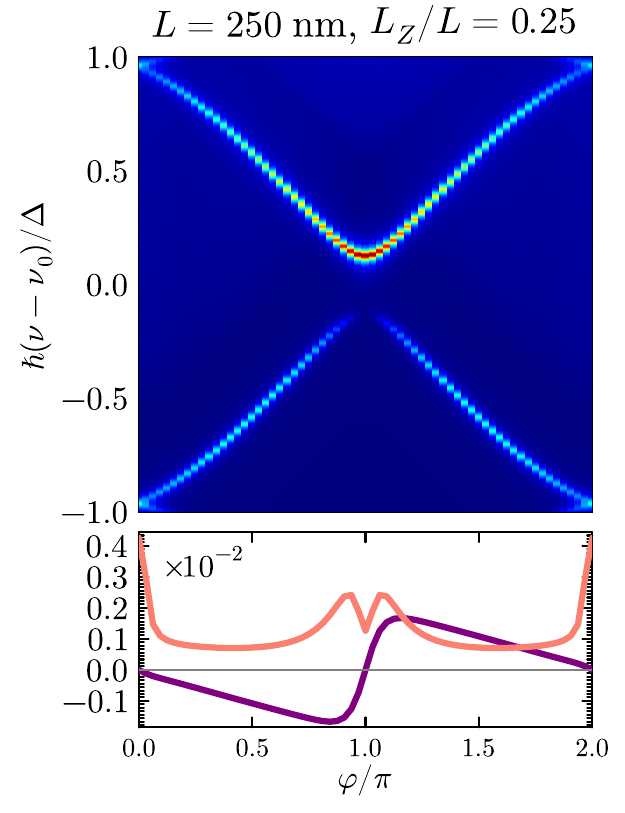}
    \\
    \includegraphics[width=5.85cm]{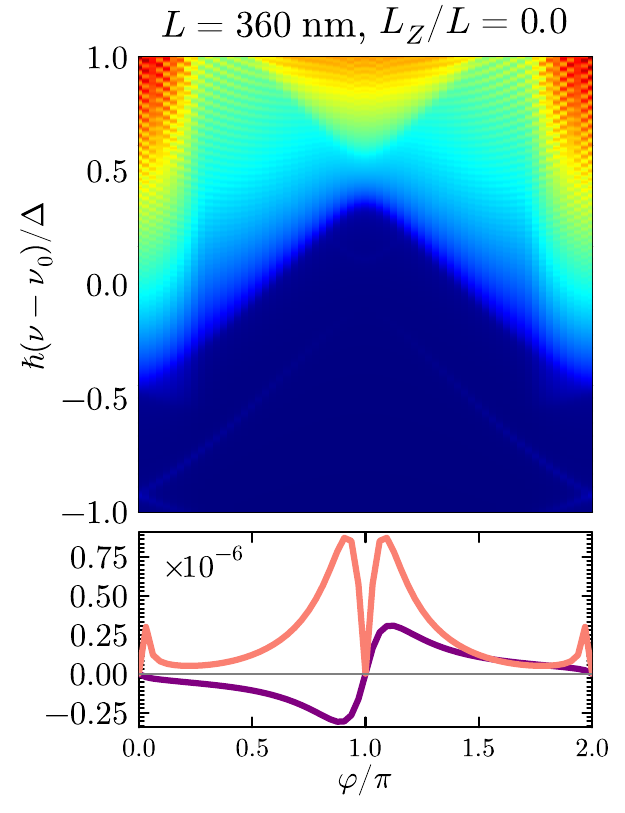}
    \includegraphics[width=5.85cm]{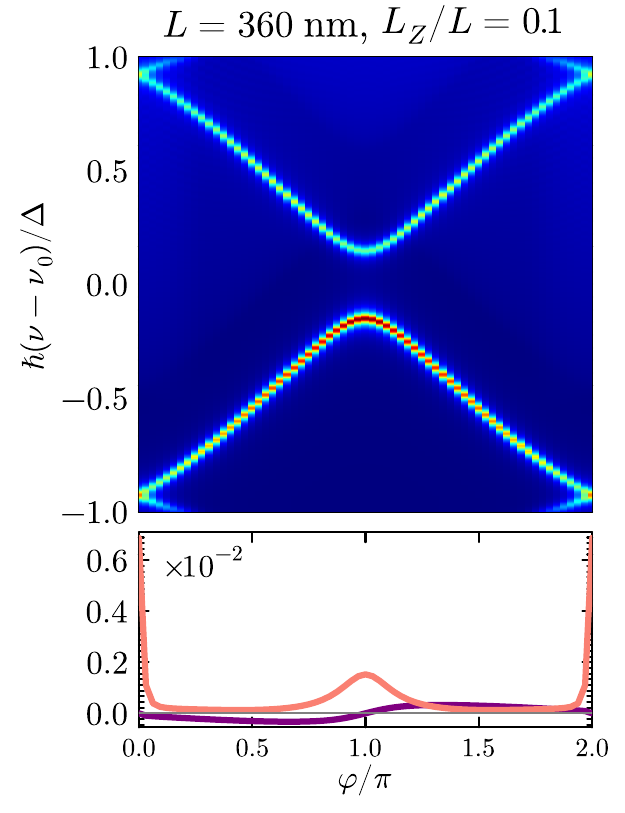}
    \includegraphics[width=5.85cm]{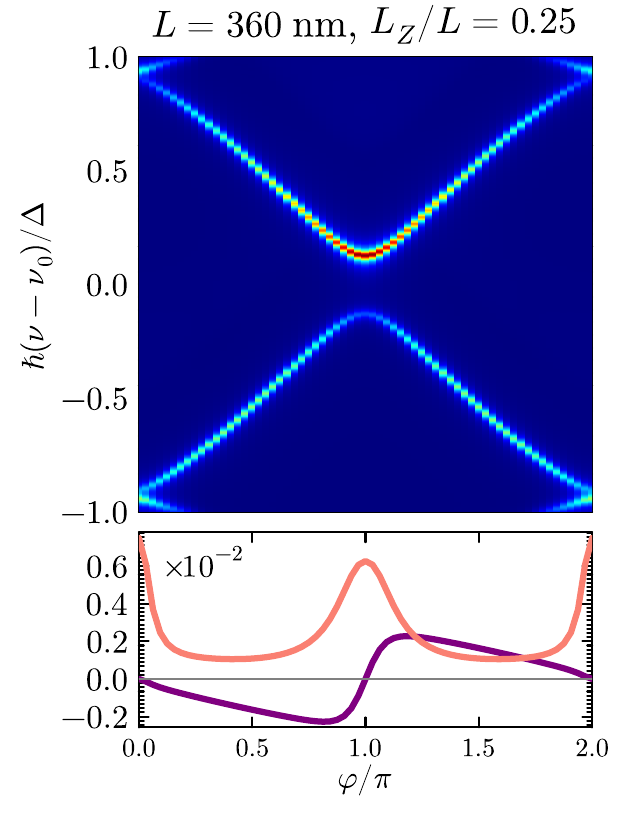}
    \\
    \includegraphics[width=5.85cm]{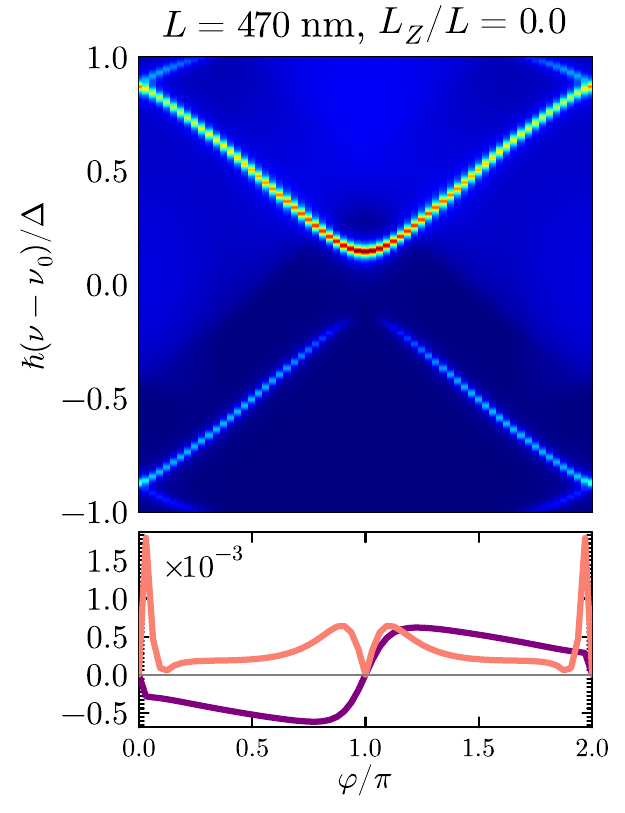}
    \includegraphics[width=5.85cm]{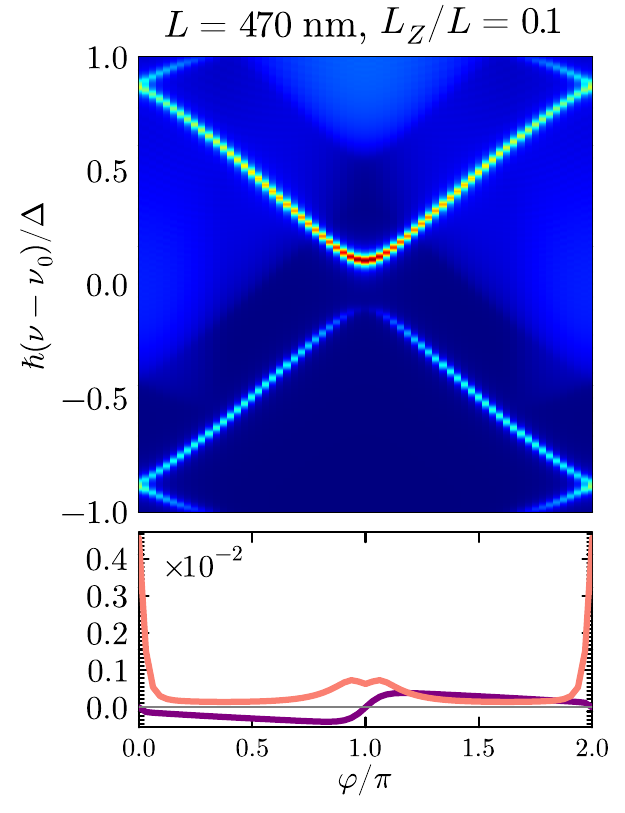}
    \includegraphics[width=5.85cm]{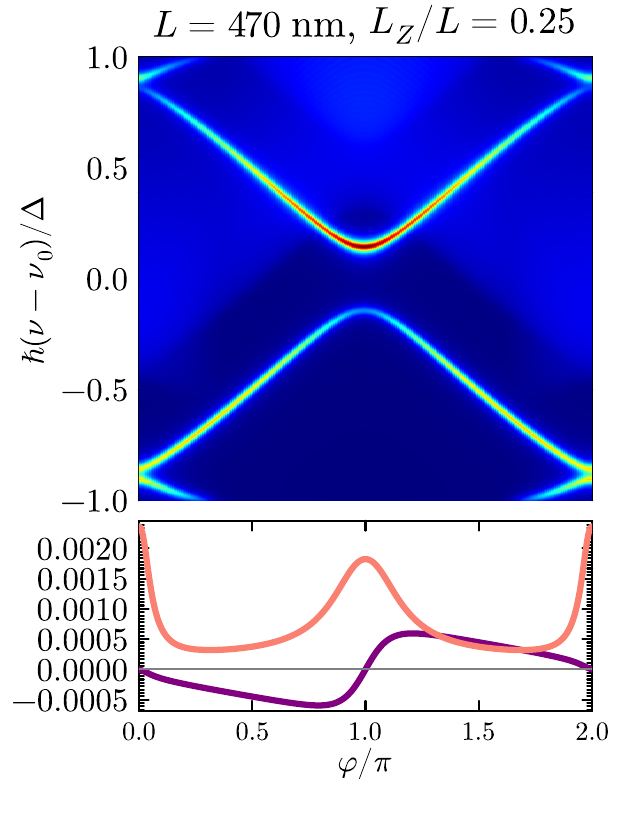}
    \caption{Absorption spectra and transduction amplitudes versus phase, for the indicated junction length $L$ and barrier position $L_Z$. The bottom right panel is a reproduction of Figs.~\ref{fig:absorption} and \ref{fig:transduction_amplitudes} of the main text. Other parameters same as in  the main text. Note the absorption color scale is not shared across panels. \label{fig:extra_params_1}}
\end{figure*}

\begin{figure*}
    \centering
    \includegraphics[width=5.85cm]{SI_figures/unified_L470nm_Zp2_kf35nm_etap02_SingleBarrierX0p00.pdf}
    \includegraphics[width=5.85cm]{SI_figures/unified_L470nm_Zp2_kf35nm_etap02_SingleBarrierX0p10.pdf}
    \includegraphics[width=5.85cm]{SI_figures/unified_L470nm_Zp2_kf35nm_etap02_SingleBarrierX0p25.pdf}
    \\
    \includegraphics[width=5.85cm]{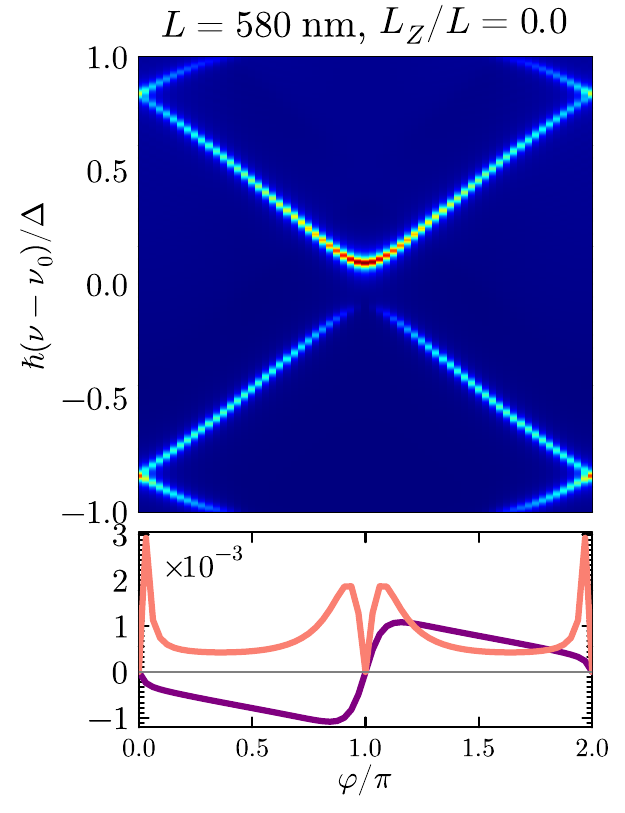}
    \includegraphics[width=5.85cm]{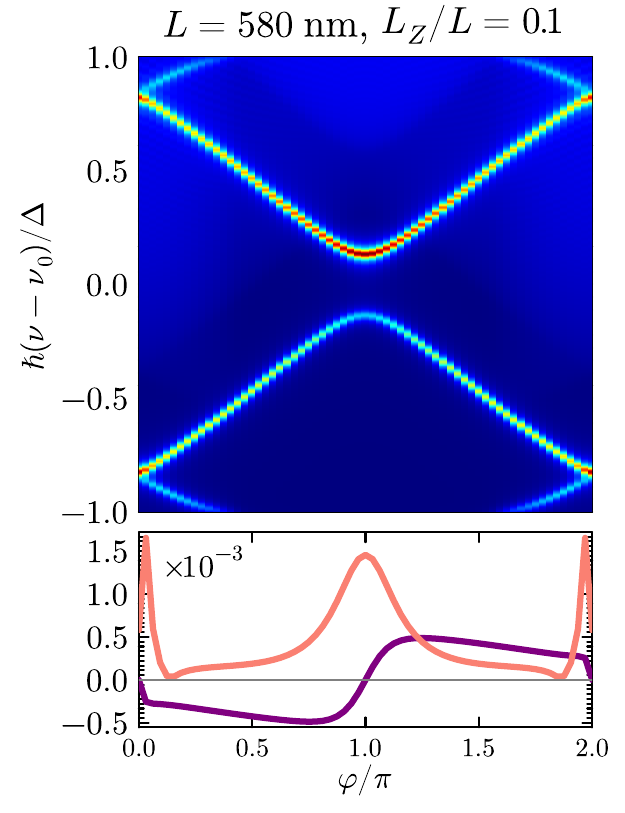}
    \includegraphics[width=5.85cm]{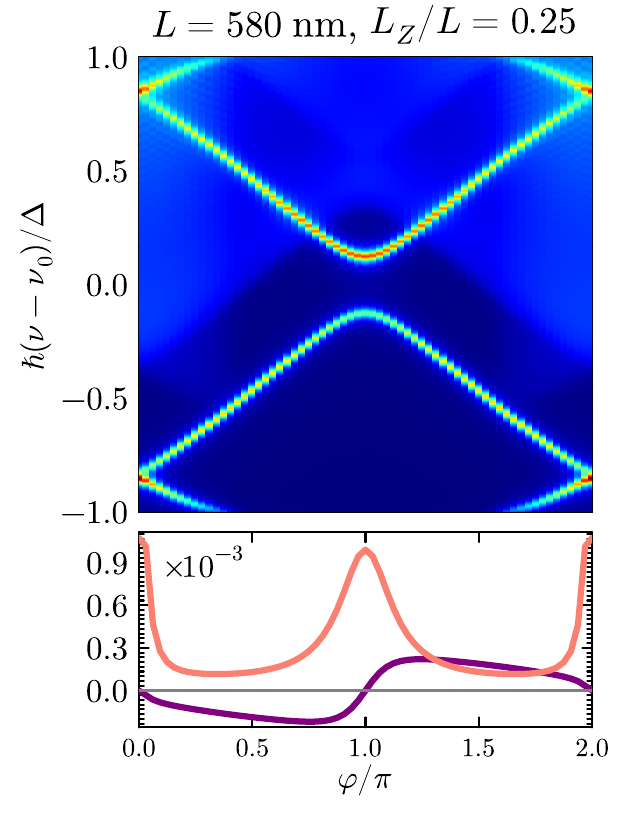}
    \\
    \includegraphics[width=5.85cm]{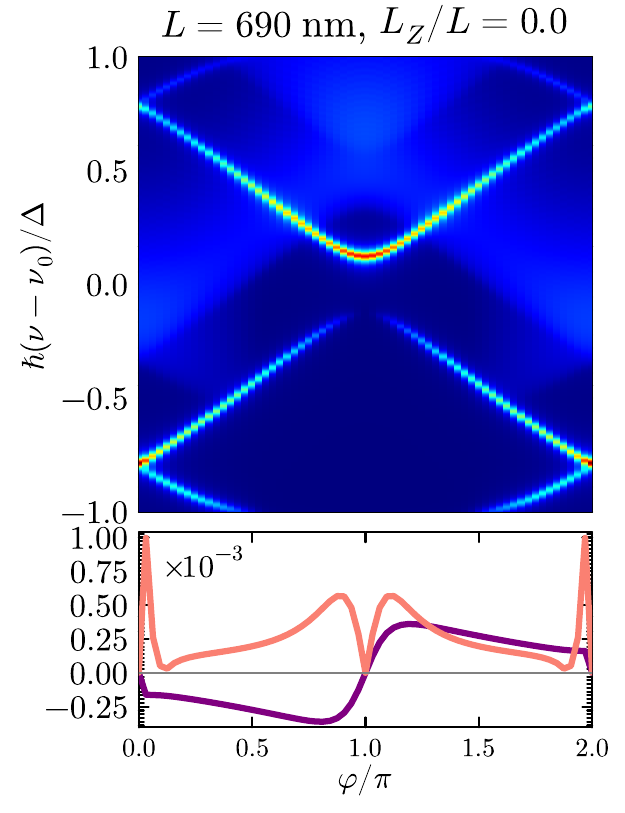}
    \includegraphics[width=5.85cm]{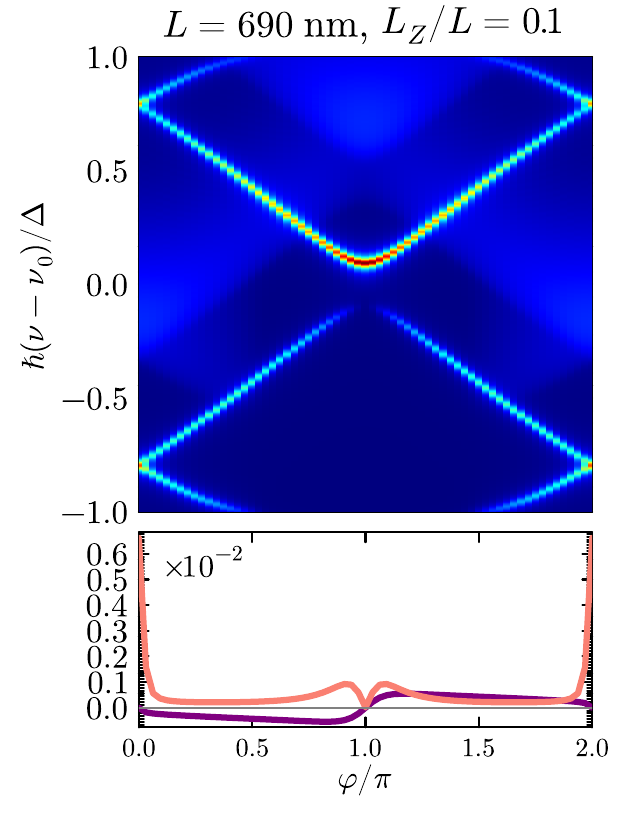}
    \includegraphics[width=5.85cm]{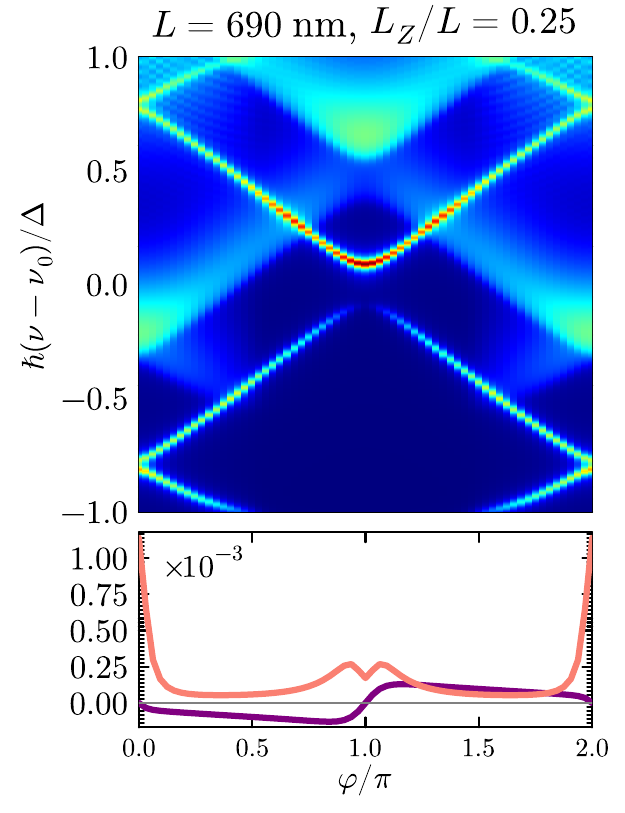}
    \caption{Continuation of Fig.~\ref{fig:extra_params_1}. \label{fig:extra_params_2}}
\end{figure*}

\bibliography{MyLibrary}